\documentclass[11pt,a4paper,english]{article}

\newcommand{\contentfolder}{./}
\def\arxivVersion{True}
\def\inlineFigures{True}

\usepackage{etoolbox}

\newcommand{\IfArxiv}[1]{ \expandafter\ifstrequal\expandafter{\arxivVersion}{True}{#1}{}}
\IfArxiv{
    \typeout{!!! arXiv Version}
    \def\inlineFigures{True}
}

\newcommand{\IfJournal}[1]{\expandafter\ifstrequal\expandafter{\arxivVersion}{True}{}{#1}}
\IfJournal{
    \typeout{!!! Journal Version}
}

\newcommand{\IfInline}[1]{\expandafter\ifstrequal\expandafter{\inlineFigures}{True}{#1}{}}
\IfInline{
    \typeout{!!! Inline figures}
    \IfJournal{
        \newcommand{\BeginFigure}{\begin{figure*}}
        \newcommand{\EndFigure}{\end{figure*}}
        \newcommand{\BeginTable}{\begin{table*}}
        \newcommand{\EndTable}{\end{table*}}
    }
    \IfArxiv{
        \newcommand{\BeginFigure}{\begin{figure}}
        \newcommand{\EndFigure}{\end{figure}}
        \newcommand{\BeginTable}{\begin{table}}
        \newcommand{\EndTable}{\end{table}}
    }
}
\newcommand{\IfNotInline}[1]{\expandafter\ifstrequal\expandafter{\inlineFigures}{True}{}{#1}}
\IfNotInline{
    \typeout{!!! Figures at end: \inlineFigures}
    \newcommand{\BeginFigure}{\begin{figure}}
    \newcommand{\EndFigure}{\end{figure}}
    \newcommand{\BeginTable}{\begin{table}}
    \newcommand{\EndTable}{\end{table}}
}

\usepackage[T1]{fontenc}

\IfArxiv{
    
    \usepackage{mathpazo}

}

\usepackage{amsmath}
\usepackage{accents}
\usepackage{amssymb, amsthm}

\IfArxiv{

    \usepackage[latin9]{inputenc}
    \usepackage[english]{babel}

    \usepackage[unicode=true,pdfusetitle,
    bookmarks=true,bookmarksnumbered=false,bookmarksopen=false,
    breaklinks=false,pdfborder={0 0 1},backref=false,colorlinks=false]
    {hyperref}

    \usepackage{array} 
    \usepackage{mathrsfs}
    \usepackage{subfigure}

    \usepackage{geometry} 

    \setlength{\topmargin}{0in}
    \setlength{\headheight}{0in}
    \setlength{\headsep}{0.0in}
    \setlength{\textheight}{8.85in}
    \setlength{\oddsidemargin}{0in}
    \setlength{\evensidemargin}{0in}
    \setlength{\textwidth}{6.5in}

    \usepackage{fancyhdr} 
    \pagestyle{fancy} 
    \lhead{}\chead{}\rhead{}
    \lfoot{}\cfoot{\thepage}\rfoot{}

    \newtheorem{definition}{Definition}

    \let\olddefinition\definition
    \renewcommand{\definition}{\olddefinition\normalfont}

}

\IfArxiv{
    \usepackage{stmaryrd}
}
\IfJournal{
    \newcommand{\llbracket}{\lsem}
    \newcommand{\rrbracket}{\rsem}
}

\usepackage{booktabs} 
\usepackage{esint}
\usepackage{breakurl}

\usepackage{paralist} 
\usepackage{verbatim} 
\usepackage{graphicx}
\usepackage{multirow}

\usepackage{braket}

\usepackage{dsfont}
\newcommand{\ident}{\mathds{1}}

\usepackage{hhtensor}

\renewcommand{\det}{\operatorname{det}}

\def\complex{\mathbb{C}}

\def\I{\ident} 
\def\dd{\mathrm{d}}

\def\st{\mathrm{st}}
\def\GP{\operatorname{GP}}

\def\E{\mathcal{E}}
\def\H{\mathcal{H}}

\newcommand\Lin{\operatorname{L}}

\newcommand\T{\operatorname{T}}

\newcommand\cnotgt{\ensuremath{\textsc{cnot}}}

\newcommand\Hil{\mathcal{H}}
\newcommand\defeq{:=} 

\makeatletter
\def\Decl@Mn@Delim#1#2#3#4{%
  \if\relax\noexpand#1%
    \let#1\undefined
  \fi
  \DeclareMathDelimiter{#1}{#2}{#3}{#4}{#3}{#4}}
\def\Decl@Mn@Open#1#2#3{\Decl@Mn@Delim{#1}{\mathopen}{#2}{#3}}
\def\Decl@Mn@Close#1#2#3{\Decl@Mn@Delim{#1}{\mathclose}{#2}{#3}}
\DeclareFontFamily{OMX}{MnSymbolE}{}
\DeclareFontShape{OMX}{MnSymbolE}{m}{n}{
    <-6>  MnSymbolE5
   <6-7>  MnSymbolE6
   <7-8>  MnSymbolE7
   <8-9>  MnSymbolE8
   <9-10> MnSymbolE9
  <10-12> MnSymbolE10
  <12->   MnSymbolE12}{}
\DeclareFontShape{OMX}{MnSymbolE}{b}{n}{
    <-6>  MnSymbolE-Bold5
   <6-7>  MnSymbolE-Bold6
   <7-8>  MnSymbolE-Bold7
   <8-9>  MnSymbolE-Bold8
   <9-10> MnSymbolE-Bold9
  <10-12> MnSymbolE-Bold10
  <12->   MnSymbolE-Bold12}{}
\DeclareSymbolFont{mnsymbols}  {OMX}{MnSymbolE}{m}{n}
\Decl@Mn@Open {\lsem}               {mnsymbols}{'102}
\Decl@Mn@Close{\rsem}               {mnsymbols}{'107}
\Decl@Mn@Open {\llangle}            {mnsymbols}{'164}
\Decl@Mn@Close{\rrangle}            {mnsymbols}{'171}
\makeatother

\newlength{\dhatheight}
\newcommand{\hhatDS}[1]{%
    \settoheight{\dhatheight}{\ensuremath{\hat{#1}}}%
    \addtolength{\dhatheight}{-0.25ex}%
    \hat{\vphantom{\rule{1pt}{\dhatheight}}%
    \smash{\hat{#1}}}}
\newcommand{\hhatTS}[1]{%
    \settoheight{\dhatheight}{\ensuremath{\hat{#1}}}%
    \addtolength{\dhatheight}{-0.25ex}%
    \hat{\vphantom{\rule{1pt}{\dhatheight}}%
    \smash{\hat{#1}}}}    
\newcommand{\hhatS}[1]{%
    \settoheight{\dhatheight}{\ensuremath{\scriptstyle{\hat{#1}}}}%
    \addtolength{\dhatheight}{-0.175ex}%
    \hat{\vphantom{\rule{1pt}{\dhatheight}}%
    \smash{\hat{#1}}}}
\newcommand{\hhatSS}[1]{%
    \settoheight{\dhatheight}{\ensuremath{\scriptscriptstyle{\hat{#1}}}}%
    \addtolength{\dhatheight}{-0.07ex}%
    \hat{\vphantom{\rule{1pt}{\dhatheight}}%
    \smash{\hat{#1}}}}
\newcommand{\hhat}[1]{\mathchoice{\hhatDS{#1}}{\hhatTS{#1}}{\hhatS{#1}}{\hhatSS{#1}}}

\def\Sket#1{\left|#1\right\rrangle}

\usepackage{authblk}

\title{Tractable Simulation of Error Correction with Honest Approximations to Realistic Fault Models}
\author[1,2]{Daniel Puzzuoli}
\author[2,3]{Christopher Granade}
\author[2,3]{Holger Haas}
\author[4]{Ben Criger}
\author[5]{\authorcr~Easwar Magesan} 
\author[2,6,7]{David Cory}
\affil[1]{Department of Applied Mathematics, University of Waterloo, Waterloo, ON, Canada}
\affil[2]{Institute for Quantum Computing, Waterloo, ON, Canada}
\affil[3]{Department of Physics and Astronomy, University of Waterloo, Waterloo, ON, Canada}
\affil[4]{Institute for Quantum Information, RWTH Aachen University, D-52056 Aachen, Germany}
\affil[5]{IBM TJ Watson Research Center, Yorktown Heights, NY, United States}
\affil[6]{Department of Chemistry, University of Waterloo, Waterloo, ON, Canada}
\affil[7]{Perimeter Institute, Waterloo, ON, Canada}

\begin{document}
\date{}
\maketitle

\begin{abstract}

In previous work, we proposed a method for leveraging efficient classical simulation algorithms to aid in the analysis of large-scale fault tolerant circuits implemented on hypothetical quantum information processors.
Here, we extend those results by numerically studying the efficacy of this proposal as a tool for understanding the performance of an error-correction gadget implemented with fault models derived from physical simulations. Our approach is to approximate the arbitrary error maps that arise from realistic physical models with errors that are amenable to a particular classical simulation algorithm in an ``honest'' way; that is, such that we do not underestimate the faults introduced by our physical models.
In all cases, our approximations provide an ``honest representation'' of the performance of the circuit composed of the original errors. This numerical evidence supports the use of our method as a way to understand the feasibility of an implementation of quantum information processing given a characterization of the underlying physical processes in experimentally accessible examples.

\end{abstract}


\section{Introduction}
\label{sec:intro}

Quantum computing is the theory and practice of using systems which exhibit the properties of quantum mechanics to store and process information, allowing certain computational problems to be solved with greater speed than any known classical algorithm \cite{shor_polynomial-time_1997,grover_fast_1996,wiebe_quantum_2012}.
This work addresses a gap in efforts towards developing a practical quantum computer,
namely that between our understanding of the physics of existing small quantum systems, and methods for reasoning about the performance of large fault-tolerant circuits. We provide a concrete algorithm that is tractable given currently available classical computational resources, and that enables reasoning about the performance of large-scale quantum information processors using experimental evidence in small systems. Our method accomplishes this task by extending the applicability of simulation algorithms for limited error models with \emph{honest} approximations of errors outside these models. This is of immediate importance given that several modalities have been proposed for the development of large-scale quantum information processors, including quantum dots \cite{jones_layered_2012} and superconducting qubits \cite{devoret_superconducting_2013}.

The development and implementation of fault-tolerant
quantum error correction (QEC) is a key milestone towards the
development of large-scale quantum computation.
Therefore, understanding the thresholds, overhead and resource requirements
of fault-tolerance methods is a critical step
towards evaluating the feasibility of proposals for large-scale
quantum information processing.
For some such fault-tolerance schemes, we can
analytically find thresholds on the acceptable errors,
such that for any error of a weaker rate than the threshold,
we can arbitrarily reduce the logical error rate.
However, analytic thresholds have not
yet been proven for fault-tolerance proposals based on topological properties,
such as surface-code based implementations \cite{fowler_high_2008}.
For these proposals, numerical simulations are performed based on
restricted error models that admit efficient classical simulation
algorithms. Connecting these numerically simulated thresholds
to models based on the physics of a device is a pressing concern
in the development and appraisal of quantum information
processing (QIP) proposals.

In recent years, proof-of-principle experiments have provided a better understanding
of the physics underlying candidate QIP devices
by implementing and fully characterizing quantum algorithms on small systems \cite{weinstein_quantum_2004}.
Such characterization techniques, however, are based on quantum process tomography \cite{chuang_prescription_1997} and are thus exponentially expensive in the size of the system.
Though techniques exist to improve the characterization of quantum systems by using
classical \cite{granade_robust_2012} and quantum \cite{wiebe_hamiltonian_2013}
simulation resources, 
the exponential cost is inevitable if we demand that the system be characterized in terms
of a quantum channel, due to the number of parameters that must be estimated.
Thus, we cannot practically hope to directly characterize intermediate or large systems
to demonstrate feasibility of QIP proposals.

A more attractive option, then, is to apply \emph{models} of noise derived
from experimental characterization of small instances of, or from simulations of the physics
underlying, a proposed device. Such realistic noise models are not
in general efficient to simulate, however, such that the question naturally arises:
how does one leverage the efficient simulation algorithms for limited noise models to
reason about the performance of large quantum information processors, using the \emph{realistic} physical models afforded by small experimental examples?
Crucially, we demand that we extrapolate small models to simulations of large quantum information processors
in an \emph{honest} fashion; that is, such that errors are only ever exaggerated and are never underestimated.

Doing so would enable the analysis of the performance of circuits implemented on a particular device, and would in turn provide insight into whether an implementation would likely succeed using a given fault-tolerance scheme. Additionally, an honest model provides a useful tool for other tasks, such as understanding resource costs of a device that is already well below the threshold. Potential applications such as these are motivated by already existent numerical studies of circuit performance that are not tied to any particular device. These studies depend on the existence of efficiently simulable \emph{sub-theories} of quantum mechanics.

The most commonly used sub-theory stems from the Gottesman-Knill (GK) theorem \cite{aaronson_improved_2004}, which provides an efficient classical simulation algorithm for acting Clifford gates and Pauli measurements on stabilizer states. Coupling this with Monte Carlo (MC) techniques, stabilizer circuits with faults modeled as the probabilistic application of stabilizer circuit elements can be simulated\IfJournal{ efficiently. While this report works exclusively with error models useful for GK-MC simulation, the general method described is independent of the set of efficiently simulable channels, making it possibly useful within the context of other efficient simulation results, such as Wigner function simulation \cite{veitch_negative_2012}, matchcircuit simulation \cite{valiant_quantum_2002}, quantum normalizer circuit simulation \cite{van_den_nest_efficient_2012,bermejo_vega_classical_2012}, or the non-adaptive strong-simulation algorithm for Clifford circuits \cite{jozsa_classical_2013}.}\IfArxiv{efficiently.\footnote{While this report works exclusively with error models useful for Gottesman-Knill and Monte-Carlo simulation, the general method described is independent of the set of efficiently simulable channels, making it possibly useful within the context of other efficient simulation results, such as Wigner function simulation \cite{veitch_negative_2012}, matchcircuit simulation \cite{valiant_quantum_2002}, quantum normalizer circuit simulation \cite{van_den_nest_efficient_2012,bermejo_vega_classical_2012}, or the non-adaptive strong-simulation algorithm for Clifford circuits \cite{jozsa_classical_2013}.} }GK-MC simulations have been used to numerically estimate threshold error rates in topological codes \cite{fowler_high_2008, bravyi_subsystem_2012, DBLP:journals/qic/WangFSH10}, and have been augmented with Sequential Monte-Carlo techniques to enable reasoning about overhead required even when well below threshold \cite{bravyi_simulation_2013}. Given the existing applications of GK-MC simulation, and the potential utility of other efficient simulation algorithms, it is highly desirable to use these techniques to aid in the understanding of real device performance. Understanding how to do this could also provide insights into how to compare the performance of experimentally realized gates to analytically derived thresholds for limited error models.

The challenge of applying efficient simulation techniques to realistic systems lies in the simple fact that errors for physical systems will, in general, not fit into one of the efficient simulation formalisms. In \cite{magesan_modeling_2013}, we proposed a solution to this problem via the concept of \emph{honest} error approximations. The idea is to replace an arbitrary error $\E$ with a channel $\Lambda$ from a restricted class $S$, such that $\Lambda$ honestly represents the tendency of $\E$ to preserve or distort quantum information. If $S$ is a subset of efficiently simulable channels, $\Lambda$ will be an efficient, honest estimate of the error induced by $\E$. While we work exclusively with our definition of honesty, we stress the generality of this proposal.

We define \emph{honesty} using the operationally motivated concept of \emph{distinguishability}. Let $\| \cdot \|_1$ denote the Schatten $1$-norm \cite{watrous_course_notes_2011}. For two quantum states $\rho_0$ and $\rho_1$, the quantity
\begin{align}\frac{1}{2} + \frac{1}{4} \| \rho_0 - \rho_1 \|_1,\end{align}
is the optimal success probability of inferring the value of a bit $\alpha \in \{0, 1\}$, drawn with uniform probability, given the state $\rho_\alpha$, using a single measurement \cite{fuchs_cryptographic_1997}. The error that a noise map $\E$ induces on a state $\rho$ is then quantified as $\| \rho - \E(\rho)\|_1$, which we term the input-output (IO) distinguishability. This provides a strong operationally motivated definition of error; given a $50\%$ chance that $\E$ acts on $\rho$, the IO distinguishability tells us what the optimal probability of ``noticing'' its action is. Using this quantification of error, the following optimization problem for finding an honest approximation to a given noise map $\E$ was given in \cite{magesan_modeling_2013}:

\begin{align}
	\text{Minimize: }& \| \Lambda - \E \|_\diamond \\
	\text{Subject to: }& \text{for every pure state } \rho,\\
	 &\| \rho - \Lambda(\rho) \|_1 \geq \| \rho - \E(\rho) \|_1, \label{eqn:honesty}
\end{align}
where $\Lambda$ ranges over some desirable subset of quantum channels (eg. Pauli channels), and $\| \cdot \|_\diamond$ is the diamond norm, which provides an analogue to distinguishability for channels \cite{kitaev_classical_2002}. 

The constraint of the above problem encodes our definition of honesty; if the constraint is satisfied for two maps $\Lambda$ and $\E$, we say that $\Lambda$ is an ``honest representation'' of $\E$. The main result of \cite{magesan_modeling_2013} is an easy-to-compute, state-independent condition on the maps $\Lambda$ and $\E$ which is sufficient to ensure honesty in the qubit case. For higher dimensions, the property ensures something similar to honesty, where the Schatten $1$-norm is replaced by the $2$-norm, though, to date, all instances that we have generated have also been found to satisfy Equation (\ref{eqn:honesty}) when tested using random pure states. We use this simpler condition in an optimization problem detailed in Appendix \ref{app:approx-dnorm} to numerically find honest approximations in this work. Details on the actual implementation of the problem, as well as a definition of the diamond norm, can also be found in Appendix \ref{app:approx-dnorm}. We note that work similar to \cite{magesan_modeling_2013} has highlighted the utility of efficiently simulable measurements for modeling non-unital errors \cite{gutierrez_approximation_2013}.

While our method ensures that the individual gate errors are honestly represented, its utility as a means for determining overall circuit performance depends on the preservation of the honesty of the approximations under composition and tensor\IfJournal{ products. (For brevity, we will abuse terminology slightly by using the word ``composition'' to refer to both direct composition and tensor products of maps.)}\IfArxiv{products.\footnote{For brevity, we will abuse terminology slightly by using the word ``composition'' to refer to both direct composition and tensor products of maps.} }That is, we are concerned with the degree to which a composition of honest approximations of gate elements is itself an honest approximation of the circuit composed of the individual gates in question.

In this paper, we numerically demonstrate that honesty is preserved under composition for some well-motivated physical models when used in a typical QEC circuit, providing evidence for the general applicability of our method. The circuit we simulate is a stabilizer circuit, and thus we approximate gate errors as probabilistic applications of stabilizer circuit elements, to make it efficiently simulable using GK-MC. Specifically, we consider approximating errors as Pauli channels and as mixed-Clifford channels (probabilistic application of Pauli gates, and Clifford gates respectively). Our analysis is carried out within a generic simulation schema, described in detail in the next section, which emulates analysis of a physical system's ability to perform a QIP task, as per \cite{jones_layered_2012, tomita_comparison_2013}. This schema serves as a demonstration of how we imagine our approximation method can be used, and tests its value within such a use case. We find that, in all tested cases, our approximations compose well; that is, the approximated circuits honestly represent the performance of the originals.

Additionally, we include performance statistics for ``Pauli twirling'', which is another way of generating a Pauli channel from an arbitrary error, within our simulation schema. In \cite{magesan_modeling_2013}, we showed that Pauli twirled approximations can underestimate the error induced by the original channel (as per our definition), and thus may not be useful for circuit performance evaluation. In \cite{geller_efficient_2013}, Geller and Zhou ask if the Pauli twirled approximations are ``sufficiently good.'' The motivation for their work is that the Pauli twirled approximation is efficiently computable in the dimension of the system, given a full description of a map, and that twirled channels are in general easier to estimate experimentally than the original \cite{emerson_symmetrized_2007,silva_scalable_2008, lopez_progress_2010}. This is in contrast to our method, which is not necessarily efficient to compute in the dimension. They argue that, while the Pauli twirled approximations sometimes underestimate the failure probability for the task they consider, the amount of underestimation is small. Our simulations, in conjunction with Geller and Zhou's work, demonstrate good and poor regimes of performance for the Pauli twirled approximations, in terms of circuit performance evaluation, with Geller and Zhou's error models falling into the good regime. In Appendix \ref{app:secondaryanalysis}, we identify and examine these regimes in detail, and argue that the poorly performing regime may be more representative of the types of errors typically found in experiment.

\IfInline{
\BeginFigure[t]
\begin{center}
  \includegraphics[width=0.8\linewidth]{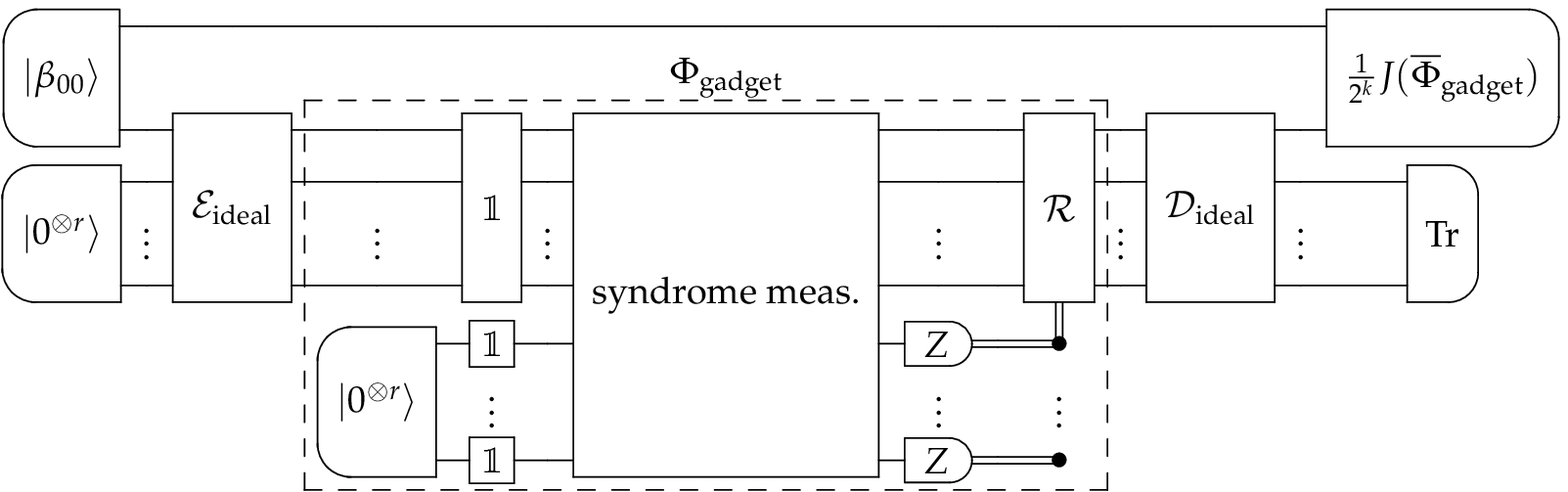}
\end{center}
  \caption{\label{fig:gadget-choi-sim-perfect} Circuit to produce a Choi state for the logical action $\overline{\Phi}_{\text{gadget}}$ of a QEC gadget acting on an $\llbracket n, k, d \rrbracket$ stabilizer code, where $r \equiv n - k$.}
\EndFigure
}

\section{Simulation Schema}
\label{sec:schema}

Our simulation schema begins with a low level physical model, and ends with a high level, efficient simulation of a QEC circuit, with our approximation method being a bridge between the two. A physical model is a description of the continuous-time dynamics of a candidate physical system for QIP, consisting of deterministic and stochastic parts to internal and control Hamiltonians, as well as dissipative open quantum-system dynamics, and includes constraints on control amplitudes. Using control techniques, a gate set is generated from a physical model, which consists of all elementary quantum logic gates required for the desired QEC circuit. Once a gate set is generated, the error on each gate is approximated using our method, yielding an \emph{honest} representation of the original gate set, which is used in an efficient simulation of the desired QEC circuit.

The circuit that we simulate, shown in Figure \ref{fig:gadget-choi-sim-perfect}, is a gadget that performs one round of error-correction in an $\llbracket n, k, d \rrbracket$ stabilizer code, which we implement using the 5-qubit perfect code (a $\llbracket 5, 1, 3 \rrbracket$ code). The circuit was chosen by balancing the desire that it be representative of standard practices, with the requirement that it be small enough to allow for fast simulation of arbitrary gate errors, thereby allowing for comparison of the efficiently simulable errors to the original. For us, a gate set consists of the gates $\{ \I, X, Y, Z, H, \cnotgt\}$, where the first four are standard single qubit Pauli gates, $H$ is the Hadamard gate, and \cnotgt~is the two qubit controlled-not\IfJournal{ gate. Note that the circuit we ultimately simulate uses only the $\I$, $H$, and $\cnotgt$ gates. The rest are included for further comparison of the gate approximations.}\IfArxiv{gate.\footnote{Note that the circuit we ultimately simulate uses only the $\I$, $H$, and $\cnotgt$ gates. The rest are included for further comparison of the gate approximations.}}

We implement this procedure for three gate sets, generated from two physical models with varying model parameters and control techniques. The physical models and control techniques are chosen to be representative of those found in experiment. Doing so allows us to encounter errors not typically considered in fault-tolerance research, despite naturally occurring in physical implementations. We emphasize however that neither the approximation method, nor the procedure for testing it given here, have been tailored for a particular outcome. The method is generic; it is independent of both the underlying physical model and gates, as well as the QEC circuit.

\section{Error Composition} \label{sec:review}

Before walking through our implementation of this scheme, we construct a simple example in which honest approximations compose dishonestly. This example demonstrates the inherent limitations of approximating an error with another error that composes in a fundamentally different way. Consider the two single-qubit maps 
\begin{align}
	\Gamma(\rho) &= U(\theta) \rho U^\dagger(\theta)\\
	\Lambda(\rho) &= (1-p)\rho + p Z \rho Z,
\end{align}
where $U(\theta)= \exp(-i\frac{\theta}{2} Z)$ and $Z$ is the Pauli $z$ operator. In \cite{magesan_modeling_2013}, it was shown that if $p = |\sin(\theta/2)|$, then $\Lambda$ and $\Gamma$ have identical IO distinguishability properties; that is, 
\begin{align}
	\| \rho - U(\theta) \rho U(\theta)^\dagger \|_1 = \| \rho - \Lambda(\rho) \|_1,
\end{align}
for all $\rho$, and is therefore $\Lambda$ an honest representation of $\Gamma$. Consider the state $\rho_+ = \ket{+}\bra{+}$, where $X \ket{+} = \ket{+}$, and $X$ is the Pauli $x$ operator, which is chosen as it is maximally sensitive to both $U(\theta)$ and $\Lambda$. One can check that 
\begin{align}
	\| \rho_+ - \Gamma(\rho_+) \|_1 = \| \rho_+ - \Lambda(\rho_+) \|_1 = 2 |\sin(\theta/2)| = 2p.
\end{align}
Now, consider the circuit composed of two applications of $\Gamma$, $C = \Gamma \circ \Gamma$, and the approximate circuit $C^{(a)} = \Lambda \circ \Lambda$, where 
\begin{align}
	C^{(a)}(\rho) = [(1-p)^2 + p^2] \rho + 2p(1-p) Z \rho Z.
\end{align} 
For $\theta \in [-\frac{\pi}{2},\frac{\pi}{2}]$, it can be checked that
\begin{align}
	\| \rho_+ - C(\rho_+) \|_1 > \|\rho_+ - C^{(a)}(\rho_+) \|_1,
\end{align}
and thus the composition of approximations is not an honest representation of the original circuit.

This example can be understood by looking at how repetitive application of these channels affects the state. Each application of $\Gamma$, a unitary error, deterministically rotates the state by a small angle, whereas $\Lambda$, a dephasing error, rotates the state by $180^\circ$, but with a small probability. In the former case, the distance that each application moves the current state remains constant, whereas in the latter case, this distance decreases exponentially, resulting in an underestimation of errors after only two applications. We encounter this situation in our simulations; one of the gate sets we consider has an identity gate error that is essentially a unitary about the $z$-axis, and the honest Pauli approximation is the dephasing channel that reproduces its IO distinguishability properties. Given this discussion and the frequency with which the gate occurs, we expected that our approximations might underestimate the overall circuit error. However, even in this case, our approximations perform as desired, providing strong numerical evidence for the value of this method in QEC circuits.

Despite these potential difficulties with error composition, it remains possible that, after a QEC protocol is applied, the resulting effective errors might compose more desirably. As an example, the first step of QEC in stabilizer codes is measuring the \emph{error syndrome}. This consists of measuring a generating set of stabilizer elements $\{Q_i \}_{i=1}^k$, which produces a $k$-bit string $b$ with $b_i = 1$ if the outcome from measuring $Q_i$ is $-1$, and $b_i = 0$ if it is $+1$. If a particular string $b$ is measured, then the system is projected onto the subspace defined by the projector 
\begin{align}
\Pi_b = 2^{-k}(\I + (-1)^{b_1}Q_1)\cdots(\I + (-1)^{b_k}Q_k).
\end{align}
For a Pauli operator $P$ and codeword $\ket{\psi}$, $\Pi_bP\ket{\psi}=0$ if $P$ does not produce syndrome $b$, and $\Pi_bP\ket{\psi} = P \ket{\psi}$ if $P$ produces syndrome $b$. Thus, indexing the Pauli operators as $\{P_i\}$, and denoting $S_b$ as the set of indices for Pauli operators that produce syndrome $b$, if a particular syndrome $b$ is measured after an error $\Lambda$, with $\chi$-matrix $\chi_{ij}$ in the Pauli basis \cite{chuang_prescription_1997}, acts on an arbitrary codeword $\ket{\psi}$, the state will be (ignoring normalization)
\begin{align*}
\begin{aligned}
	\Pi_b \Lambda(\ket{\psi}\bra{\psi}) \Pi_b &= \sum_{ij} \chi_{ij} \Pi_b P_i \ket{\psi}\bra{\psi} P_j \Pi_b\\
	&= \sum_{ij \in S_b} \chi_{ij} P_i \ket{\psi} \bra{\psi} P_j.
\end{aligned}
\end{align*}
Given this form, it is clear that if $P_i$ and $P_j$ have different syndromes, then $\chi_{ij}$ can play no part in the post-syndrome measurement state, and is therefore effectively truncated by syndrome measurement. In this way, errors become more ``incoherent'' and this, at least superficially, makes errors ``more like'' Pauli channels (which have diagonal $\chi$-matrices in the Pauli basis). Thus, whatever the form of $\Lambda$, after a correction step is enacted, the effective error may compose more like a Pauli channel. In practice, this argument may fail due to various aspects of imperfect syndrome measurement, such as limited visibility measurements, and the time it takes to perform measurement protocols like ancilla-assisted syndrome measurement.


\section{Implementing the Schema}
\subsection{Physical Models and Gate Set Generation}

We consider two physical models, PM1 and PM2. PM1 is motivated by a double quantum dot system, and PM2 represents an archetypal two level system (see Appendix \ref{app:physical-models} for a description and full details). Gate Set 1 (GS1) is built on PM1, and Gate Sets 2 and 3 (GS2 and GS3) are built on PM2 (using different model parameters). GS1 and GS2 use noise refocusing techniques \cite{gullion_new_1990,maudsley_modified_1986}, which mitigate errors induced by stochastic Hamiltonians. GS2 and GS3 implement gates via hard pulses; that is, the pulse sequences used to generate the gates are manually specified by choosing control amplitudes. Due to the complicated structure of the Hamiltonian in PM1, optimal control theory (OCT) was used to find pulse sequences that implement the gates in GS1 with high fidelity \cite{khaneja_optimal_2005,borneman_application_2010}. Every gate in a set is made to be the same length in time, as our circuit simulation proceeds in discrete time steps in which a single gate acts on every register qubit. A full description of how the gates are simulated is given in Appendix \ref{app:cumulant}.

The different combinations of physical model and control techniques give rise to different types of gate errors. A detailed account on the form of the errors for each gate set is given in Appendix \ref{app:secondaryanalysis}, as it has particular relevance within the context of that discussion. We do however wish to highlight that some gates in GS1 have largely unitary errors, resulting from the use of OCT pulse finding in gate implementation. Thus, given the discussion on error composition, GS1 provides a strong test for our method.
\subsection{Gate Set Approximations and Statistics}

For each gate set we generate three efficiently simulable approximate gate sets:
\begin{description}
	\item[Pauli twirled] The Pauli twirled errors.
	\item[Pauli] The honest Pauli channel approximation.
	\item[Clifford] The honest mixed-Clifford channel\IfJournal{ approximation. Note that only the single-qubit errors are approximated as mixed-Cliffords (we allowed the algorithm to search over all mixed-Cliffords). Due to the large number of two-qubit Cliffords, the honest Pauli approximation for the \cnotgt~gate is reused.}\IfArxiv{approximation.\footnote{Note that only the single-qubit errors are approximated as mixed-Cliffords (we allowed the algorithm to search over all mixed-Cliffords). Due to the large number of two-qubit Cliffords, the honest Pauli approximation for the \cnotgt~gate is reused.}}
\end{description}
Several metrics are used to compare how each approximation performs on individual gates. In what follows we denote the noisy implementation of some ideal operation $U_{\text{ideal}}$ by $\Lambda_{\text{Original}}$, and use $\Lambda$ as a place holder for the various approximations. The first few metrics are well known quantities.
\begin{itemize}
	\item $\chi_{00}$ \textemdash~The first entry of the $\chi$-process matrix of the error in the Pauli basis \cite{chuang_prescription_1997}. This quantity is reported due to its relation to the average gate fidelity \cite{nielsen_simple_2002}, and, for Pauli channels, is the probability that no fault occurs.
	\item $\| \Lambda - U_{\text{Ideal}} \|_\diamond$ and $\| \Lambda - \Lambda_{\text{Original}} \|_\diamond$ \textemdash~The distance of the approximation $\Lambda$ to the ideal gate and original error, respectively.
\end{itemize}

The rest stem from our definition of honesty. Using the function 
\begin{align}
	h(\Gamma, \E, \rho) \equiv \| \rho - \Gamma(\rho) \|_1 - \| \rho - \E(\rho) \|_1,
\end{align}
which we call the \emph{hedging} of the channel $\Gamma$ relative to $\E$ for the state $\rho$, the statement that $\Lambda$ honestly represents the error of $\Lambda_{\text{Original}}$ can be restated as $h(\Lambda, \Lambda_{\text{Original}}, \rho) \geq 0$ for all pure states $\rho$. We calculate three quantities related to the hedging, which we approximate by randomly sampling $N$ pure states $\{ \ket{\psi_i} \}_{i=1}^N$.
\begin{itemize}
	\item $\bar{h}(\Lambda, \Lambda_{\text{Original}})\equiv \int d\psi \: h(\Lambda, \Lambda_{\text{Original}}, \ket{\psi} \bra{\psi})$\\ $\approx \frac{1}{N} \sum_{i=1}^N h(\Lambda, \Lambda_{\text{Original}}, \ket{\psi_i} \bra{\psi_i})$ \textemdash~The average of the hedging function over pure states.
	\item $p_{\text{viol}} \approx \frac{N_{\text{viol}}}{N}$ \textemdash~The ratio of pure states $\ket{\psi}$ for which $h(\Lambda, \Lambda_{\text{Original}}, \ket{\psi} \bra{\psi}) < 0$, where $N_{\text{viol}} = \left| \left\{ \ket{\psi} \in \{\ket{\psi_i}\}_{i=1}^N : h(\Lambda, \Lambda_{\text{Original}}, \ket{\psi} \bra{\psi}) < 0\right\}\right|$.
\end{itemize}
Table \ref{table:identity-stats} presents the statistics for the identity gate from each gate set, using $N=10^6$ uniformly sampled pure states. See Appendix \ref{app:gate-statistics} for tables containing statistics on all gates.

\begin{table}[h!] 
\begin{center}
    \caption{Statistics for the various approximations of the identity gate for each gate set, approximated from $N=10^6$ random pure states.}
\label{table:identity-stats} 
\begin{tabular}{ccllll} \toprule
\multicolumn{2}{c}{Set/Statistics} & $\hphantom{-}$Original& $\hphantom{-}$Pauli twirled& $\hphantom{-}$Pauli& $\hphantom{-}$Clifford \\ \midrule 
\multirow{6}{*}{GS1} & $\chi_{00}$ & $\hphantom{-}0.999994$ & $\hphantom{-}0.999994$ & $\hphantom{-}0.997618$ & $\hphantom{-}0.998314$ \\ 
 			& $\| \Lambda - U_{\text{Ideal}} \|_{\diamond}$ & $\hphantom{-}4.76\times 10^{-3}$ & $\hphantom{-}1.20\times 10^{-5}$ & $\hphantom{-}4.76\times 10^{-3}$ & $\hphantom{-}4.77\times 10^{-3}$ \\ 
 			& $\| \Lambda - \Lambda_{\text{Original}} \|_{\diamond}$ &  & $\hphantom{-}4.76\times 10^{-3}$ & $\hphantom{-}6.73\times 10^{-3}$ & $\hphantom{-}3.64\times 10^{-3}$ \\ 
 			& $\overline{h}$ &  & $-3.73\times 10^{-3}$ & $\hphantom{-}1.14\times 10^{-7}$ & $\hphantom{-}1.64\times 10^{-6}$ \\ 
 			& $p_{\text{viol}}$ &  & $\hphantom{-}1.$ & $\hphantom{-}0.$ & $\hphantom{-}0.$ \\ \midrule 
\multirow{6}{*}{GS2} & $\chi_{00}$ & $\hphantom{-}0.999087$ & $\hphantom{-}0.999087$ & $\hphantom{-}0.999085$ & $\hphantom{-}0.999086$ \\ 
 			& $\| \Lambda - U_{\text{Ideal}} \|_{\diamond}$ & $\hphantom{-}1.83\times 10^{-3}$ & $\hphantom{-}1.83\times 10^{-3}$ & $\hphantom{-}1.83\times 10^{-3}$ & $\hphantom{-}1.83\times 10^{-3}$ \\ 
 			& $\| \Lambda - \Lambda_{\text{Original}} \|_{\diamond}$ &  & $\hphantom{-}2.48\times 10^{-5}$ & $\hphantom{-}2.50\times 10^{-5}$ & $\hphantom{-}2.06\times 10^{-6}$ \\ 
 			& $\overline{h}$ &  & $-1.63\times 10^{-7}$ & $\hphantom{-}2.65\times 10^{-6}$ & $\hphantom{-}8.34\times 10^{-7}$ \\ 
 			& $p_{\text{viol}}$ &  & $\hphantom{-}0.49861$ & $\hphantom{-}0.$ & $\hphantom{-}0.$ \\  \midrule 
\multirow{6}{*}{GS3} & $\chi_{00}$ & $\hphantom{-}0.998751$ & $\hphantom{-}0.998751$ & $\hphantom{-}0.996501$ & $\hphantom{-}0.996501$ \\ 
 			& $\| \Lambda - U_{\text{Ideal}} \|_{\diamond}$ & $\hphantom{-}4.99\times 10^{-3}$ & $\hphantom{-}2.50\times 10^{-3}$ & $\hphantom{-}7.00\times 10^{-3}$ & $\hphantom{-}7.00\times 10^{-3}$ \\ 
 			& $\| \Lambda - \Lambda_{\text{Original}} \|_{\diamond}$ &  & $\hphantom{-}2.50\times 10^{-3}$ & $\hphantom{-}5.28\times 10^{-3}$ & $\hphantom{-}5.28\times 10^{-3}$ \\ 
 			& $\overline{h}$ &  & $-1.03\times 10^{-3}$ & $\hphantom{-}1.91\times 10^{-3}$ & $\hphantom{-}1.91\times 10^{-3}$ \\ 
 			& $p_{\text{viol}}$ &  & $\hphantom{-}0.74978$ & $\hphantom{-}0.$ & $\hphantom{-}0.$ \\ \bottomrule 
\end{tabular}
\end{center}
\end{table}

Looking at the various diamond norm distances between the Original, Ideal, Pauli twirled ($\Lambda_\text{PT}$), and Pauli errors ($\Lambda_\text{P}$), a simple ordering can be seen to hold for every gate:
\begin{align*}
\begin{array}{c}
	\left\| \Lambda_\text{PT} - U_\text{Ideal} \right\|_\diamond \leq \left\| \Lambda_\text{Original} - U_\text{Ideal} \right\|_\diamond \leq \left\| \Lambda_\text{P} - U_\text{Ideal} \right\|_\diamond, \\[5pt]
	\left\| \Lambda_\text{PT} - \Lambda_\text{Original} \right\|_\diamond \leq \left\| \Lambda_\text{P} - \Lambda_\text{Original} \right\|_\diamond.
\end{array}
\end{align*}
Thus, while the Pauli twirled error is always closer than the Pauli to the Original, it is also always closer to the Ideal than the Original, and is therefore a less noticeable error than the Original. The location of the Clifford approximation in the second inequality chain varies; for some gates, it is an order of magnitude closer to the Original than both the Pauli and Pauli twirled approximations, for others, it is the same distance to the Original as the Pauli, indicating that the best Pauli is also the best Clifford approximation, and for the rest, it is in between the two.

Two other important and connected observations can be made. In some cases, the $\chi_{00}$ element of the honest approximations is much lower than that of the Original, and in others it is not appreciably different. In the former cases, the Pauli twirled approximations tend to be much closer to the Ideal, and have worse hedging performance, than in the latter cases. These are demonstrations of channels with different average fidelities but similar IO distinguishability properties, and channels with identical average fidelities but very different IO distinguishability properties. This observation is connected to the different regimes of performance for Pauli twirling, as well as how ``coherent'' or ``unitary'' an error is, and is explained in detail in Appendix \ref{app:secondaryanalysis}.
\subsection{Circuit Design and Simulation Results}
\begin{figure}[h!]
  \begin{center}
  \includegraphics[scale=1]{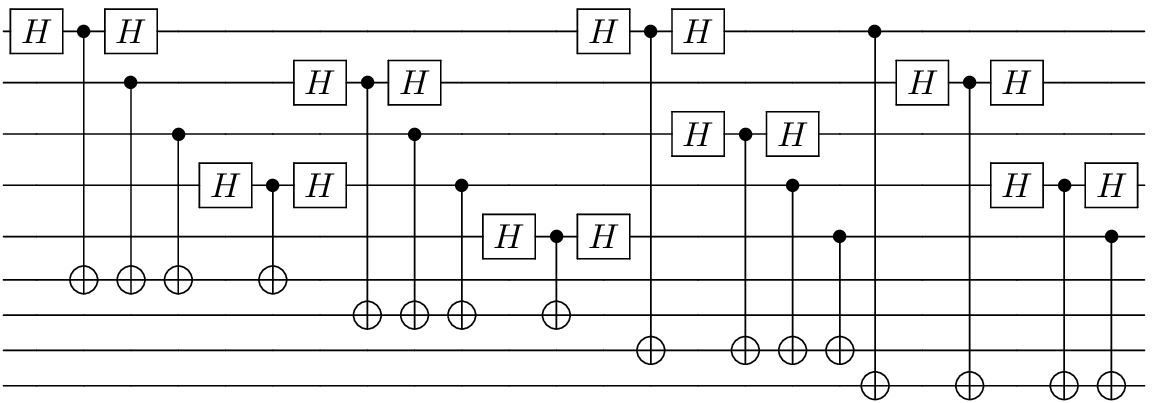}
  
  (a) Original circuit.
  \vskip0.5em
  \includegraphics[width=0.9\linewidth]{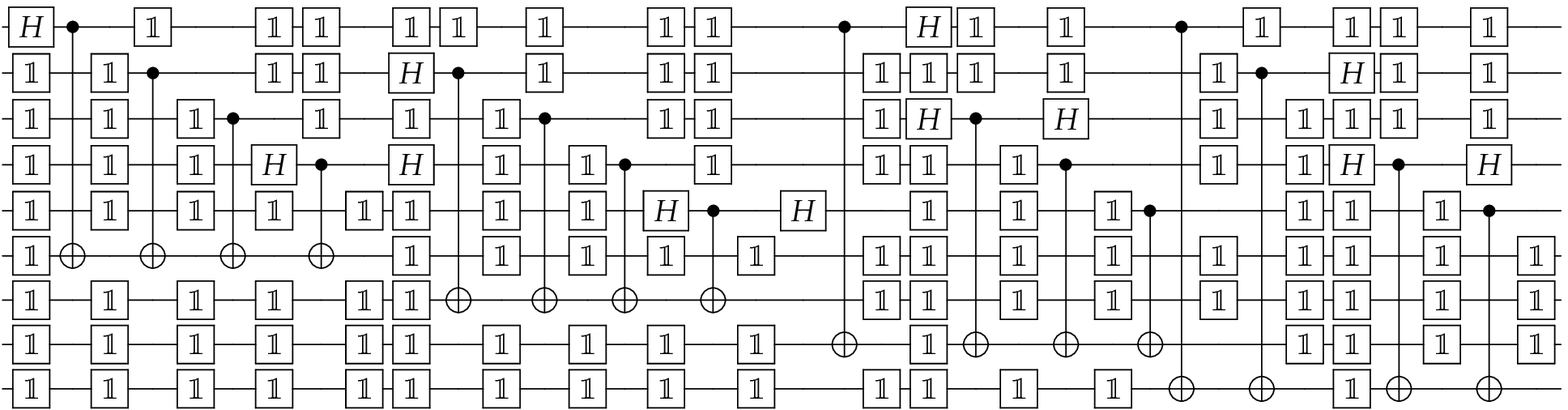}
  
  (b) Circuit with simplifications and with explicit wait locations.
  
  \end{center}
  \caption{
    \label{fig:perfect-synmeas} Syndrome measurement circuit for the five-qubit perfect code.
  }
\end{figure}

\begin{figure}[h!]
 \[
  \includegraphics[scale=1]{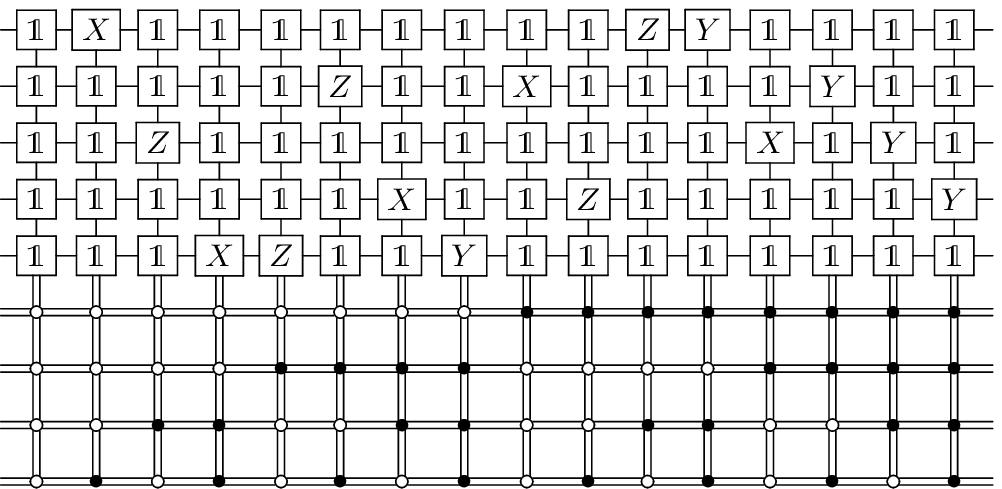}
  \]
  \caption{
    \label{fig:perfect-recovery}
    Recovery circuit for the five-qubit perfect code.
  }
\end{figure}

We simulate the gadget $\Phi_{\text{gadget}}$ that performs one round of error correction on one block of an error-correcting code, as per Figure \ref{fig:gadget-choi-sim-perfect}.
We isolate the action $\overline{\Phi}_{\text{gadget}}$ of this gadget on the encoded state by preceding and following it with perfect encoding and decoding operations, $\mathcal{E}_{\text{ideal}}$ and $\mathcal{D}_{\text{ideal}}$. The circuit is simulated by computing its Choi state; for a code that encodes $k$ logical qubits into $n$ physical qubits, we take the state
\begin{align}
	\ket{\beta_{00}} = \frac{1}{\sqrt{2^k}} \sum_{i=1}^{2^k} \ket{i} \otimes \ket{i}, \label{eqn:bell-state}
\end{align}
where $\{ \ket{i} \}_{i=1}^{2^k}$ is an orthonormal basis for the $k$-qubit Hilbert space, and compute 
\begin{align}
	\left(C\otimes \I_{L(\complex^k)}\right)\left(\ket{\beta_{00}} \bra{\beta_{00}}\right) = \frac{1}{2^k} J(C),
\end{align}
where $C$ represents the entire circuit and $J(C)$ is the Choi-Jamio\l{}kowski matrix for the circuit. (This simulation method is chosen due to the efficiency with which maps of the form we consider can be applied to states. See Appendix \ref{app:computationaltools-circuitsim} for details.) Due to the perfect encoding and decoding operations, $J(C) = J(\overline{\Phi}_{\text{gadget}})$. Explicitly, the circuit performs the following operations:
\begin{enumerate}
	\item The state $\ket{\beta_{00}}$ (Equation \ref{eqn:bell-state}) is prepared, half of which is perfectly encoded into an $\llbracket n, k, d \rrbracket$ stabilizer code.
	\item The gadget $\Phi_{\text{gadget}}$ is applied to the encoded physical qubits, consisting of:
		\begin{enumerate}[i)]
			\item One imperfect wait location on all of the data qubits. This is a placeholder for possible non-trivial operations in gadgets meant to perform logical operations.
			\item Simultaneously, a register of ancillas for ancilla-assisted syndrome measurement is prepared. An imperfect identity operation acts on each ancilla to represent imperfect ancilla preparation.
			\item Imperfect ancilla-assisted syndrome measurement is performed (see Figure \ref{fig:perfect-synmeas}). Measurement of the physical ancillas is taken to be perfect, with errors represented by identity gates that precede the measurement.
		\end{enumerate}
	\item A perfect recovery operation is performed by classical feed-forward of syndrome measurement details (see Figure \ref{fig:perfect-recovery}).
	\item Once $\Phi_{\text{gadget}}$ is done, the resultant state is then perfectly decoded and the physical ancillas are discarded.
\end{enumerate}

The recovery operation is chosen to be perfect as, in practice, it isn't always necessary to physically perform the recovery; errors can be tracked and taken into account when further operations are performed on the block~\cite{Knill2005}.

We implement this simulation schema using the 5-qubit perfect code, with the syndrome measurement and recovery circuits designed by the Python package QuaEC \cite{quaec}. Note that we choose to perform the syndrome measurement in a non-fault-tolerant way, as a fault-tolerant gadget for a code with $n$ physical qubits would require at least $2n$ ancilla qubits for the Steane or Knill fault-tolerant error correction (FTEC) gadgets, or strictly more than $\sum_i \text{wt}(S_i)=16$ ancillae for the Shor FTEC gadget. Thus, at least 10 ancillae are needed for the perfect code, requiring simulation of at least 16 qubits, putting us outside the range of quickly simulable circuits with arbitrary errors.

\begin{figure}[h!] 
\begin{center}
  \includegraphics{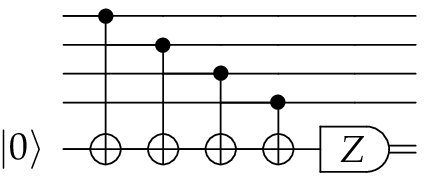}
\end{center}
  \caption{\label{fig:fowler-schedule} Circuit to measure the stabilizer generator $Z^{\otimes 4}$, proposed by Fowler et al \cite{fowler_high_2008} for use in surface codes.}
\end{figure}

With 4 stabilizer generators, this code requires 4 ancillas for encoding and 4 for syndrome measurement using the circuit shown in Figure \ref{fig:perfect-synmeas}. Any redundant Hadamard gates have been removed. The recovery operation is shown in Figure \ref{fig:perfect-recovery}. This type of non-fault tolerant syndrome measurement is similar to gadgets proposed for use in topological QEC codes, such as the surface code. In particular, the syndrome measurement gadget used by Fowler et al \cite{fowler_high_2008}, shown in Figure \ref{fig:fowler-schedule}, relies on \cnotgt~gates between each data qubit in the support of a stabilizer generator and a common ancilla qubit. We emphasize that we are not concerned with \emph{absolute} circuit performance. Rather, the task at hand is the comparison of \emph{relative} performance of efficiently simulable error approximations, so it suffices that this circuit contains all of the typical elements and procedures for QEC, regardless of fault-tolerance.

Table \ref{table:sim-results} gives the simulation statistics for $\overline{\Phi}_{\text{gadget}}$, again using $N=10^6$ sample pure states to compute the hedging parameters. For each gate set, the Pauli and Clifford approximations compose well; the approximated circuit honestly represents the error of the original. This is especially encouraging for GS1, given its unitary identity error. For the Pauli twirled errors, we see that in GS1 they fail the honesty condition for every tested pure state. For GS2, they fail the honesty condition for just over half of the pure states tested, but by an arguably small degree. Interestingly, for GS3, the Pauli twirled approximations provide an honest representation of the circuit performance.

\begin{table}[h!] %
\caption{Statistics for $\overline{\Phi}_{\text{gadget}}$ using $N=10^6$ randomly sampled pure states.}\label{table:sim-results}
\begin{center}
\begin{tabular}{ccllll} \toprule 
\multicolumn{2}{c}{Set/Statistics} & $\hphantom{-}$Original& $\hphantom{-}$Pauli twirled& $\hphantom{-}$Pauli& $\hphantom{-}$Clifford \\ \midrule 
\multirow{6}{*}{GS1} & $\chi_{00}$ & $\hphantom{-}0.999964$ & $\hphantom{-}0.999964$ & $\hphantom{-}0.985820$ & $\hphantom{-}0.989930$ \\ 
 			& $\| \Lambda - U_{\text{Ideal}} \|_{\diamond}$ & $\hphantom{-}4.76\times 10^{-3}$ & $\hphantom{-}7.28\times 10^{-5}$ & $\hphantom{-}2.84\times 10^{-2}$ & $\hphantom{-}2.04\times 10^{-2}$ \\ 
 			& $\| \Lambda - \Lambda_{\text{Original}} \|_{\diamond}$ &  & $\hphantom{-}4.76\times 10^{-3}$ & $\hphantom{-}2.87\times 10^{-2}$ & $\hphantom{-}2.01\times 10^{-2}$ \\ 
 			& $\overline{h}$ &  & $-3.69\times 10^{-3}$ & $\hphantom{-}1.85\times 10^{-2}$ & $\hphantom{-}1.23\times 10^{-2}$ \\ 
 			& $p_{\text{viol}}$ &  & $\hphantom{-}1.$ & $\hphantom{-}0.$ & $\hphantom{-}0.$ \\  \midrule
\multirow{6}{*}{GS2} & $\chi_{00}$ & $\hphantom{-}0.991372$ & $\hphantom{-}0.991372$ & $\hphantom{-}0.991355$ & $\hphantom{-}0.991367$ \\ 
 			& $\| \Lambda - U_{\text{Ideal}} \|_{\diamond}$ & $\hphantom{-}1.73\times 10^{-2}$ & $\hphantom{-}1.73\times 10^{-2}$ & $\hphantom{-}1.73\times 10^{-2}$ & $\hphantom{-}1.73\times 10^{-2}$ \\ 
 			& $\| \Lambda - \Lambda_{\text{Original}} \|_{\diamond}$ &  & $\hphantom{-}2.45\times 10^{-5}$ & $\hphantom{-}4.29\times 10^{-5}$ & $\hphantom{-}1.14\times 10^{-5}$ \\ 
 			& $\overline{h}$ &  & $-1.63\times 10^{-8}$ & $\hphantom{-}2.24\times 10^{-5}$ & $\hphantom{-}7.66\times 10^{-6}$ \\ 
 			& $p_{\text{viol}}$ &  & $\hphantom{-}0.55566$ & $\hphantom{-}0.$ & $\hphantom{-}0.$ \\   \midrule 
\multirow{6}{*}{GS3} & $\chi_{00}$ & $\hphantom{-}0.992495$ & $\hphantom{-}0.987594$ & $\hphantom{-}0.969499$ & $\hphantom{-}0.969499$ \\ 
 			& $\| \Lambda - U_{\text{Ideal}} \|_{\diamond}$ & $\hphantom{-}1.51\times 10^{-2}$ & $\hphantom{-}2.48\times 10^{-2}$ & $\hphantom{-}6.10\times 10^{-2}$ & $\hphantom{-}6.10\times 10^{-2}$ \\ 
 			& $\| \Lambda - \Lambda_{\text{Original}} \|_{\diamond}$ &  & $\hphantom{-}1.03\times 10^{-2}$ & $\hphantom{-}4.60\times 10^{-2}$ & $\hphantom{-}4.60\times 10^{-2}$ \\ 
 			& $\overline{h}$ &  & $\hphantom{-}6.36\times 10^{-3}$ & $\hphantom{-}3.04\times 10^{-2}$ & $\hphantom{-}3.04\times 10^{-2}$ \\ 
 			& $p_{\text{viol}}$ &  & $\hphantom{-}0.$ & $\hphantom{-}0.$ & $\hphantom{-}0.$ \\   \bottomrule 
\end{tabular}
\end{center}
\end{table}

\section{Conclusion}
\label{sec:conclusion}

In all examined cases, the honest approximations led to honest representations of circuit performance, providing confidence in our method as a tool for evaluating the performance of typical QEC circuits with realistic gate errors. By starting from continuous-time physical models, and building gates using common control techniques, we tested our method against errors typical of those found in experiment. The details of the physical models and control techniques were not tailor-made for any desired outcome. The strongest test of our method came from errors with strong unitary parts, arising from OCT designed pulses, a regime not typically considered in fault-tolerance research. Additionally, our results, in conjunction with the recent work by Geller and Zhou \cite{geller_efficient_2013}, demonstrate two regimes of performance for Pauli twirled error approximations. In one regime, their performance can be considered ``sufficiently good'', while in the other, Pauli twirling results in systematic underestimation of the IO distinguishability notion of error (see Appendix \ref{app:secondaryanalysis}).

Our work is motivated by the desire for the simulations to be pessimistic. We want to be reasonably assured that, if the simulation with the approximated errors performs well according to some metric, then the actual implementation will perform well also. Currently, experimental implementations of QIP are limited to small system sizes. Extrapolating their performance to hypothetical large-scale systems requires caution. Quantum processors will require constant application of error-correction protocols like the one we consider, and it is imaginable that in large systems, consisting of hundreds of qubits or more, even a small underestimation of the effect of physical-level errors may dramatically compound, resulting in false expectations of over-all performance.

This work provides hope that our method can be used as a tool for extrapolating performance from small systems in an honest way. This is key to understanding the feasibility of various proposals for large-scale quantum devices, which in turn will aid in planning the way forward for feasible experimental implementations of QIP.

\IfJournal{\begin{acknowledgments}}
The authors would like to thank Christopher Wood, Ian Hincks, Troy Borneman, John Watrous, and Joseph Emerson for helpful discussions, and acknowledge support from NSERC, CIFAR, and the Government of Ontario through OGS. This work was partially supported by the Intelligence Advanced Research Projects Activity (IARPA) via Department of Interior National Business Center, Contract No. DIIPC20166. The US Government is authorized to reproduce and distribute reprints for Governmental purposes notwithstanding any copyright annotation thereon. Disclaimer: The views and conclusions contained herein are those of the authors and should not be interpreted as necessarily representing the official policies or endorsements, either expressed or implied, of IARPA, DoI/NBC, or the US Government.
\IfJournal{\end{acknowledgments}}



\appendix
\section{Methods}
\label{app:methods}

\subsection{Channel Approximation and Diamond Norm Computation} \label{app:approx-dnorm}

To measure the distance between two maps $\Lambda$ and $\E$, we use the diamond norm distance $\| \Lambda - \E \|_\diamond$, which provides an analogue to distinguishability for channels \cite{kitaev_classical_2002}. For any map $\Phi : \Lin(\H_1) \rightarrow \Lin(\H_2)$ (mapping the linear operators acting on one Hilbert space to the linear operators acting on another), the diamond norm can be defined as:
\begin{align}
	\left\| \Phi \right\|_\diamond \equiv \max \left\{ \left\| (\Phi\otimes \I_{\Lin(\H_1)})(X)\right\|_1 : X \in \Lin(\H_1 \otimes \H_1), \|X \|_1 \leq 1\right\},
\end{align}
where $\I_{\Lin(\H_1)}$ is the identity channel acting on $\Lin(\H_1)$. Given a set of channels $S$, to find an honest approximation of an error map $\E$, we seek to minimize $\| \Lambda - \E \|_\diamond$ for $\Lambda \in S$ such that $\Lambda$ satisfies the constraint given in \cite{magesan_modeling_2013}. To implement the optimization, we require that $S$ has a parameterization. For our purposes, it is natural to consider sets of the form $S = \{ \sum_{i=1}^n p_i \Lambda_i : p_i \geq 0 \text{ and } \sum_{i=1}^n p_i = 1 \}$. Given this, we have the following optimization problem:

\begin{align}
	\begin{aligned}
	&\text{input: Finite set of channels } \{ \Lambda_i \}_{i=1}^n \text{and channel }\E\\
	&\text{minimize: } f(p_1, ..., p_n) = \left\| \sum_{i=1}^n p_i \Lambda_i - \E \right\|_\diamond\\
	&\text{s.t. } \sum_{i=1}^n p_i = 1, p_i \geq 0, \text{ and } A \geq 0\text{, where}\\
	&\quad \; \; A = (\I - M_{\Lambda})^T(\I - M_{\Lambda}) - (\I - M_{\E})^T(\I - M_{\E}) +\\
	&\quad \; \; (\| \vec{t}_{\Lambda} \|_2^2 -\| \vec{t}_{\E} \|_2^2 - 2 \| (\I - M_{\Lambda})^T \vec{t}_{\Lambda} - (\I - M_{\E})^T \vec{t}_{\E}\|_2)\I,
 	\end{aligned}
\end{align}
where $M_\Lambda = \sum_{i=1}^n p_i M_i$, $\vec{t}_\Lambda = \sum_{i=1}^n p_i \vec{t}_i$, $(M_i, \vec{t}_i)$ is the Bloch representation of channel $\Lambda_i$, $(M_\E, \vec{t}_\E)$ is the Bloch representation of $\E$, and $\I$ is the identity matrix of appropriate size. Note that the constraint given here is slightly more general than the one in \cite{magesan_modeling_2013}, as it includes the possibility for $S$ to contain non-unital channels. We note again that if a qubit map $\Lambda$ satisfies the constraint in the above optimization problem, then it is an \emph{honest} representation of the error $\E$, according to the definition in the introduction. For higher dimensional channels, the constraint is sufficient to ensure something similar to honesty, where the Schatten $1$-norm is replaced by the Schatten $2$-norm, in which case the approximation $\Lambda$ is still guaranteed to be somehow globally worse than $\E$, though not in an operationally motivated way. To date, however, all higher dimensional approximations that we have generated using this algorithm have also been found to be honest when tested using random pure states.

We implement the approximation optimization problem in MATLAB using the built in fmincon function. The diamond norm is computed using a semi-definite program given by Watrous \cite{watrous_diamond_norm} and implemented using the CVX package \cite{cvx}. Linear constraints ensure that the vector $(p_1, ..., p_n)$ is a probability vector and non-linear constraints check that the eigenvalues of the matrix $A$ are non-negative with a tolerance of $10^{-15}$. The SQP algorithm is used for the optimization. Due to the non-convexity of this problem it is necessary to run many local solvers and then choose the best result. This is done using the MultiStart function which instantiates the local solver many times over randomly chosen starting points that satisfy the constraints. For each approximation, we used $72$ starting points.

\subsection{Cumulant Simulation} \label{app:cumulant}
To simulate quantum logic gates using our noise models, a method for the simulation of stochastic quantum evolution is required. This has been considered in the context of analyzing the fidelity with which decoherence-free subsystems can be implemented \cite{cappellaro_principles_2006}. In that case, the cumulant expansion \cite{kubo_generalized_1962, vankampen_cumulant_1_1973, vankampen_cumulant_2_1973} was applied to model the effects of stochastic dynamics on a quantum system.

Following that approach, we will consider that, conditioned on a particular realization of noise, our system evolves according to the Liouville-von Neumann equation
\begin{equation}
  \label{eq:liouville-vonneumann}
  \frac{\partial}{\partial t} \rho(t) = -i [H(t), \rho(t)] + D[\rho(t)],
\end{equation}
where $\rho(t)$ is the density operator describing our system at time $t$, $H$ is the Hamiltonian of the system, and where $D \in \T(\Hil)$ is a linear transformation describing the decoherence of the system. We assume that $H(t)$ can be decomposed into deterministic and stochastic parts,
\begin{equation}
  H(t) = H_{\det}(t) + H_{\st}(t).
\end{equation}
We then further decompose $H_{\st}(t)$ such that all of the stochasticity is encapsulated in a set of scalar-valued functions $\{\omega_1(t), \dots, \omega_k(t)\}$. Thus,
\begin{equation}
  H_{\st}(t) = \sum_i \omega_i(t) A_i(t)
\end{equation}
for some set of deterministic operator-valued functions $\{A_i(t)\}$.

To analyze the dissipation transformation $D$, we assume that it can be written in Lindblad form,
\begin{equation}
  D[\rho(t)] = \sum_i L_i \rho(t) L_i^{\dagger} - \frac{1}{2} \{L_i^\dagger L_i, \rho(t)\},
\end{equation}
where $\{L_i\}$ are called the Lindblad operators of the system.

Both the Liouvillian operator $\mathcal{L}[\rho(t)] \defeq [H, \rho(t)]$ and the dissipation operator $D$ act linearly on density operators, and thus may be represented by superoperators $\hhat{\mathcal{L}}, \hhat{D} \in \Lin(\Lin(\Hil))$, where $\Lin(\Hil)$ marks the set of all linear operators acting on Hilbert space $\Hil$. Using the isomorphism that $\Lin(\Hil) \cong \Hil\otimes\Hil$, we shall use the column-stacking basis for $\Hil\otimes\Hil$, such that $\Sket{\ket{i}\bra{j}} = \ket{j}\otimes\ket{i}$. Therefore one can rewrite Equation \eqref{eq:liouville-vonneumann} as
\begin{equation}
   \frac{\partial}{\partial t} \Sket{\rho(t)} = \left( -i[ \hhat{\mathcal{L}}_\text{det}(t) + \hhat{\mathcal{L}}_\text{st}(t)] + \hhat{D} \right) \Sket{\rho(t)} .
\end{equation}
Now we go to the rotating frame of the deterministic superoperator $\hhat{\mathcal{L}}_\text{det}(t)$, i.e. we define a unitary $\mathcal{U}(t) = \T \exp\left(-i \int_0^t \hhat{\mathcal{L}}_\text{det}(t') \dd t'\right)$ such that 
\begin{equation}
   \label{eq:rot-frame-liouville-vonneumann}
   \frac{\partial}{\partial t} \Sket{\tilde{\rho}(t)} = \left( -i~\mathcal{U}^\dagger(t) \hhat{\mathcal{L}}_\text{st}(t) \mathcal{U}(t) + \mathcal{U}^\dagger(t) \hhat{D} \mathcal{U}(t) \right) \Sket{\tilde{\rho}(t)} ,
\end{equation}
where $\Sket{\tilde{\rho}(t)} = \mathcal{U}^\dagger(t) \Sket{\rho(t)} $.

The formal solution to Equation \eqref{eq:rot-frame-liouville-vonneumann}, then, for a single realization of the trajectories $\{\omega(t)\}$ is given by
\begin{equation}
  \Sket{\tilde{\rho}(t)} = \T \exp\left(-i \int_0^t \hhat{G}(t') \dd t'\right) \Sket{\rho(0)},
\end{equation}
with $\hhat{G}(t) \defeq ~\mathcal{U}^\dagger(t) \hhat{\mathcal{L}}_\text{st}(t) \mathcal{U}(t) + i~ \mathcal{U}^\dagger(t) \hhat{D} \mathcal{U}(t)$.

For our purposes, we are interested in the average evolution $\hhat{S}$ over the ensemble of control trajectories,
\begin{equation}
  \hhat{S}(t) = \left\langle \T \exp\left(-i \int_0^t \hhat{G}(t') \dd t'\right) \right\rangle.
\end{equation}
The cumulant expansion gives us that $\hhat{S}(t) = \exp(\hhat{K}(t))$, where
\begin{align}
  \label{eq:cumulant-generator}
  \hhat{K}(t) & = \sum_{n=1}^{\infty} \frac{(-it)^n}{n!} K_n = -it \hhat{K}_1 - \frac{t^2}{2} \hhat{K}_2 + \cdots, \\
  \hhat{K}_1  & = \frac{1}{t}      \int_0^t \dd t_1 \left\langle \hhat{G}(t_1) \right\rangle, \label{eq:K1} \\
  \hhat{K}_2  & = \frac{1}{t^2} \T \int_0^t \dd t_1 \int_0^t \dd t_2 \left\langle \hhat{G}(t_1) \hhat{G}(t_2) \right\rangle - \hhat{K}^2_1.
\end{align}
To simplify this, we assume that each control parameter $\omega_i(t)$ is a trajectory of a stationary zero-mean process (The zero-mean assumption technically isn't an assumption; the mean of each random process can be absorbed into the deterministic part of the Hamiltonian.) That is, that $\vec{\omega} \sim \GP(0, \matr{\phi})$, where $\matr{\phi}$ is the matrix-valued autocorrelation function for $\vec{\omega}(t)$, such that $\phi_{i,j} (t_1 - t_2) = \langle \omega_i(t_1) \omega_j(t_2) \rangle$.

Then $\hhat{K}_1$ becomes simply
\begin{equation}
  \label{eq:K1-autocorrelation}
  \hhat{K}_1 = \frac{i}{t} \int_0^t \dd t_1~\mathcal{U}^\dagger(t_1) \hhat{D} \mathcal{U}(t_1) ,
\end{equation}
whereas we can then rewrite $\hhat{K}_2$ in terms of the autocorrelation function,
\begin{align}
  \label{eq:K2-autocorrelation}
  \hhat{K}_2 &= \frac{2}{t^2} \int_0^t \dd t_1 \int_0^{t_1} \dd t_2
     \sum_{i,j=1}^k \phi_{i,j} (t_1 - t_2) \mathcal{U}^\dagger(t_1) \hhat{A}_i(t_1) \mathcal{U}(t_1) \mathcal{U}^\dagger(t_2) \hhat{A}_j(t_2) \mathcal{U}(t_2) \\
     &- \frac{2}{t^2} \int_0^t \dd t_1 \int_0^{t_1} \dd t_2 ~ \mathcal{U}^\dagger(t_1) \hhat{D} \mathcal{U}(t_1) \mathcal{U}^\dagger(t_2) \hhat{D} \mathcal{U}(t_2) - \hhat{K}^2_1,
\end{align}
where $\hhat{A}_i(t) = -A_i^*(t) \otimes \I + \I\otimes A_i(t)$. In this way, we note that the cumulant expansion generalizes the Magnus expansion so as to account for stochastically-varying fields. The motivation for using cumulants instead of expanding the time-ordered exponential in terms of moments of the stochastic process stems from the fact that cumulant averages enter in the exponential, reducing the risk of truncation artefacts. 

To numerically simulate the gate action, we discretize $\hhat{\mathcal{L}}_\text{det}(t')$ along our gate length $t$ at $N$ points, with equal time intervals $\Delta t$ between these points, i.e. we evaluate $\{\hhat{\mathcal{L}}_\text{det}(m \Delta t)\}$, with $m=1,...,N$ while $t=N \Delta t$. Next we approximate $\mathcal{U}(n \Delta)$ by
\begin{equation}
\label{eq:Uapprox}
\mathcal{U}(n \Delta) \approx \exp \left( -i \hhat{\mathcal{L}}_\text{det}(n \Delta t) \Delta t \right)~...~\exp \left( -i \hhat{\mathcal{L}}_\text{det}(\Delta t) \Delta t \right) \exp \left( -i \hhat{\mathcal{L}}_\text{det}(0) \Delta t \right) .
\end{equation}
Finally, we turn turn the integral in line \eqref{eq:K1-autocorrelation} into a sum
\begin{equation}
   \hhat{K}_1 \approx \frac{i}{N} \sum_{n=0}^{N-1} ~\mathcal{U}^\dagger(n \Delta t) \hhat{D} \mathcal{U}(n \Delta t) ,
\end{equation}
and the double integral in line \eqref{eq:K2-autocorrelation} into a double sum
\begin{align}
  \label{eq:K2-approx}
  \begin{aligned}
  \hhat{K}_2 &\approx \frac{1}{N^2} \sum_{n=0}^{N-1} \sum_{i,j=1}^k \phi_{i,j} (0) \mathcal{U}^\dagger(n \Delta t) \hhat{A}_i(n \Delta t) \mathcal{U}(n \Delta t) \mathcal{U}^\dagger(n \Delta t) \hhat{A}_j(n \Delta t) \mathcal{U}(n \Delta t) \\
  &+\frac{2}{N^2} \sum_{n=1}^{N-1} \sum_{m=0}^{n-1} 
  \sum_{i,j=1}^k \phi_{i,j} ((n-m) \Delta t) \mathcal{U}^\dagger(n \Delta t) \hhat{A}_i(n \Delta t) \mathcal{U}(n \Delta t) \mathcal{U}^\dagger(m \Delta t) \hhat{A}_j(m \Delta t) \mathcal{U}(m \Delta t) \\
  &-\frac{1}{N^2} \sum_{n=0}^{N-1} \mathcal{U}^\dagger(n \Delta t) \hhat{D} \mathcal{U}(n \Delta t) \mathcal{U}^\dagger(n \Delta t) \hhat{D} \mathcal{U}(n \Delta t) \\
   &- \frac{2}{N^2} \sum_{n=1}^{N-1} \sum_{m=0}^{n-1} \mathcal{U}^\dagger(n \Delta t) \hhat{D} \mathcal{U}(n \Delta t) \mathcal{U}^\dagger(m \Delta t) \hhat{D} \mathcal{U}(m \Delta t) - \hhat{K}_1^2.
\end{aligned}
\end{align}

To simulate the gates, we truncated $\hat{K}(t)$ in Equation \eqref{eq:cumulant-generator} at second order, which can be partially justified with the following. If we have no dissipator term in Equation \eqref{eq:liouville-vonneumann}, then due to statistical independence, the $m$th order cumulant disappears if, for a set of times $\{t_1, t_2, ..., t_n\}$, any of the time gaps $|t_1-t_2|$, $|t_2-t_3|$, ...,$|t_{n-1}-t_n|$ are larger than the correlation time $\tau_c$ of the stochastic process \cite{kubo_generalized_1962}. Since cumulants at every order vanish once the gap between the set of time points exceeds $\tau_c$, then if $t \gg \tau_c$,  the $m$th order cumulant $\hhat{K}_m$ is effectively an integral over an $(m-1)$ dimensional sphere with radius $\tau_c$, integrated over $t$. Therefore, $\hhat{K}_m$ scales roughly as $\tau_c^{m-1} A^m t$, where $A$ is the maximum norm of $\hhat{A}_i(t)$. Comparing the second- and fourth-order cumulants, $\hhat{K}_2$ and $\hhat{K}_4$, reveals that $\frac{\tau_c^{3} A^4 t}{\tau_c A^2 t} = \tau_c^{2} A^2$, meaning that if $\tau_c A \ll 1$ and $t \gg \tau_c$, we have a justification for truncating the cumulant expansion at second order. For the physical models considered in this work, the dissipator terms in the Liouville-von Neumann equation were considerably smaller in their norm than the noise Hamiltonian terms, so we assume that the arguments above are still applicable.

\subsection{Circuit Simulation} \label{app:computationaltools-circuitsim}
Each gate in the gate set acts on either one or two qubits, subjecting the rest to identical, uncorrelated noise (the noisy identity gate). The action of a noisy process on a quantum register can be calculated in the Kraus representation,
\begin{equation}
\Lambda (\rho) = \sum_{j=1}^{4^n} A_j \rho A_j^{\dagger}.
\end{equation}
For a generic noisy process, using na\"ive matrix multiplication, this calculation involves $\sim 2^{5n}$ operations and requires the storage of $\sim 2^{4n}$ complex parameters. These costs can be reduced dramatically by exploiting the fact that the noise is independent and that the gate set acts identically on different qubits. Noise maps that act independently commute, and can be applied in sequence as a result:
\begin{equation}
\lambda^{\otimes n} = \lambda_1 (\lambda_2 (\ldots \lambda_n(\rho)))
\end{equation}
Thus, the amount of storage is reduced to that required to store the gate set and the current state, $\sim 2^{2n}$ complex parameters, and the number of operations required now scales as $n2^{3n}$. This can be further reduced by noting that each channel $\lambda_j$ is equivalent to the perfect identity on $n-1$ qubits, and its effect can be pre-calculated to reduce the total number of operations to $n2^{2n}$. The extension to two-qubit gates is straightforward; for further information, see \cite{QuTiP}.

\section{Physical Models and Gate Protocol Details} \label{app:physical-models}

This\IfJournal{ Supplementary Information}\IfArxiv{Appendix }describes the physical models, noise refocusing techniques, and the subsequent gate sets generated from these. It is assumed that the density matrix $\rho(t)$ describing a physical system evolves according to
\begin{equation}
  \frac{\partial}{\partial t} \rho(t) = -i \hbar [H(t),\rho(t)] 
  + \sum_i \left( L_i \rho(t) L_i^{\dagger} - \frac{1}{2} \{ L_i^{\dagger} L_i, \rho(t) \} \right),
\end{equation}
where $H(t)$ is the Hamiltonian for the system and $\{L_i\}$ is a set of Lindblad operators generating non-unitary dynamics \cite{lindblad_generators_1976,gorini_completely_1976,breuer_theory_2002}. A physical model must specify all deterministic and stochastic Hamiltonians (both internal, and control), and specify any Lindblad operators that the system is subject to.

\subsection{Physical Model 1}
Physical Model 1 is motivated by a double quantum dot physical system. A double quantum dot is a pair of electrons contained in a double potential well. The spatial and spin states of the electrons encode logical states $\ket{0}$ and $\ket{1}$,
\begin{align}
  \ket{0} &= \ket{\Phi_{11}^T} \otimes \left( \ket{\uparrow \downarrow} + \ket{\downarrow \uparrow} \right) / \sqrt{2} \label{eq:0-state} \\
  \ket{1} &= \left( a \ket{\Phi_{11}^S} + b \ket{\Phi_{02}} \right) \otimes \left( \ket{\uparrow \downarrow} - \ket{\downarrow \uparrow} \right) / \sqrt{2} \label{eq:1-state},
\end{align}
where $\ket{\Phi_{11}^S}$ and $\ket{\Phi_{11}^T}$ are symmetric and anti-symmetric spatial states with one electron in each of the potential wells, and $\ket{\Phi_{02}^S}$ is a symmetric spatial state having two electrons in one particular well.

The electron state can be controlled by varying the voltage detuning $B(t)$ and Zeeman splitting difference $A(t)$, described below.
\begin{itemize}
  \item {\bf Voltage detuning} introduces a potential energy difference $B$ between the quantum wells. $B > 0$ favours the $\ket{\Phi_{02}}$ spatial state over $\ket{\Phi_{11}^S}$ and $\ket{\Phi_{11}^T}$, because $\ket{\Phi_{02}}$ allows for both electrons to minimize their potential energy. The parameters $a$ and $b$ in Equation (\ref{eq:1-state}) are therefore $B$-dependent and given by Fermi-Dirac statistics \cite{Landau1969} such that the probability $p_{11}=|a|^2$ of having an electron with potential energy $B$ is given by $p_{11}=\frac{1}{1+e^{B/B_1 - B_2}}$, whereby
  \begin{align*}
    a(B) &= \sqrt{\frac{1}{1+e^{B/B_1 - B_2}}} \\
    b(B) &= \sqrt{\frac{1}{1+e^{-(B/B_1 - B_2)}}} .
  \end{align*}
  The detuning Hamiltonian $H(B)$ is diagonal in the spatial states $\{\ket{\Phi_{11}^S}, \ket{\Phi_{11}^T}, \ket{\Phi_{02}}\}$ and takes a form
\begin{align}H(B) = B \left( \ket{\Phi_{11}^S} \bra{\Phi_{11}^S} + \ket{\Phi_{11}^T} \bra{\Phi_{11}^T} \right) + B_0\ket{\Phi_{02}} \bra{\Phi_{02}},\end{align}
where $B_0$ is the energy eigenvalue of $\ket{\Phi_{02}}$ at zero detuning. Up to a constant identity contribution, this yields the following Hamiltonian for logical states
  \begin{equation*}
H(B) = \left(
    \begin{array}{cc}
      \bra{0} H(B) \ket{0} & \bra{0} H(B) \ket{1} \\
      \bra{1} H(B) \ket{0} & \bra{1} H(B) \ket{1} \\
    \end{array}
    \right) = \frac{1}{2} \left(
    \begin{array}{cc}
      |b|^2 (B-B_0) & 0 \\
      0 & -|b|^2 (B-B_0) \\
    \end{array}  
  \right) .
  \end{equation*}
  \item {\bf Zeeman splitting} is related to the energy difference between electron spin-up and spin-down states in the presence of an external magnetic field. A magnetic field gradient across the potential wells introduces an energy splitting $A$ between $\ket{\Phi_{11}} \otimes \ket{\uparrow \downarrow}$ and $\ket{\Phi_{11}} \otimes \ket{\downarrow \uparrow}$, where $\ket{\Phi_{11}} = \ket{\Phi_{11}^S}+\ket{\Phi_{11}^T}$, leading to a Hamiltonian\\ $H(A) = \frac{A}{2} \left( \ket{\Phi_{11}} \bra{\Phi_{11}} \otimes \ket{\uparrow \downarrow} \bra{\uparrow \downarrow} - \ket{\Phi_{11}} \bra{\Phi_{11}} \otimes \ket{\downarrow \uparrow} \bra{\downarrow \uparrow} \right)$. The resulting matrix $H(A)$ in the logical basis is
  \begin{equation*}
H(A) = \left(
    \begin{array}{cc}
      \bra{0} H(A) \ket{0} & \bra{0} H(A) \ket{1} \\
      \bra{1} H(A) \ket{0} & \bra{1} H(A) \ket{1} \\
    \end{array}
    \right) = \frac{1}{2} \left(
    \begin{array}{cc}
      0 & a A \\
      a^* A & 0 \\
    \end{array}
    \right) .
  \end{equation*}
\end{itemize}

Substituting the parameters $a(B)$ and $b(B)$ into the above expressions and summing them results in the effective logical single qubit Hamiltonian
\begin{align}
	H(t) = \frac{1}{2} \frac{A(t) + \alpha(t)}{\sqrt{1 + \exp \left( \frac{B(t)}{B_1} - B_2 \right)}} X + \frac{1}{2} \frac{B(t) - B_0 + \beta(t)}{1 + \exp \left[-\left(\frac{B(t)}{B_1} - B_2 \right) \right]} Z,\label{eq:QDHam}
\end{align}
where the parameters $\alpha(t)$ and $\beta(t)$ encapsulate the stochastic behaviour of the parameters $A(t)$ and $B(t)$. As the logical state $\ket{0}$ corresponds to the ground state of the physical Hamiltonian at zero detuning $(B=0)$ and in zero magnetic field gradient $(A=0)$, we make an assumption that relaxation acts on the logical state simply as a Lindblad operator $L = \frac{1}{2 \sqrt{T_1}}(X + i Y)$.

The parameters $A(t)$, $B(t)$, $B_i$ for $i \in \{0,1,2\}$, and  $T_1$, define the deterministic evolution of the system, with the first two representing the single-qubit controls and the latter being the $T_1$ time-constant of the system. Both control scalars are specified for intervals of length $\delta t$, over which the controls remain constant, whereas the rate of change of these control scalars between adjacent intervals is bounded with $\left|\frac{d A(t)}{d t} \right| \leq \Delta A_\text{max}$ and $\left|\frac{d B(t)}{d t}\right| \leq \Delta B_\text{max}$. Additionally, the controls are bounded by some maximum value; that is $\left| A(t) \right| \leq A_\text{max}$ and $0 \leq B(t) \leq B_\text{max}$ for some $A_\text{max},B_\text{max} \geq 0$. 

The parameters specified by $\alpha$ and $\beta$ are independent, zero-mean, stationary Gaussian processes \cite{rasmussen_gpml_2005}, such that $\langle \alpha(t) \rangle = \langle \beta(t) \rangle = 0$ and $\langle \alpha(t_1) \beta(t_2) \rangle = 0$. The auto-correlation functions are given by
\begin{align}
\begin{aligned}
	\langle \alpha(t_1)\alpha(t_2) \rangle &= \Gamma_{\alpha_1}^2~\delta(|t_1-t_2|) + \Gamma_{\alpha_2}^2 e^{-\left(\frac{|t_1 -t_2|}{\tau_1}\right)^2 +\left( \frac{|t_1 -t_2|}{\tau_2} \right)^4-\left( \frac{|t_1 -t_2|}{\tau_3} \right)^6} \\
	\langle \beta(t_1)\beta(t_2) \rangle &= \Gamma_{\beta_1}^2 + \Gamma_{\beta_2}^2~\delta(|t_1-t_2|) ,
\end{aligned}
\end{align}
where $\delta(t)$ is the Dirac delta function. Parameters labelled with the letter $\Gamma$ represent the noise strengths, and those labelled with $\tau$ represent various correlation times.

The Hamiltonian for simulating two-qubit gates is given by
\begin{align}
   \nonumber H(t) &= H^{(1)}(t) \otimes \I + \I \otimes H^{(2)}(t) + H_{zz}(t) \\
   \nonumber H_{zz}(t) &= \frac{1}{4} \frac{C(t)(1+ \gamma(t))}{\left( 1 + \exp \left[-\left(\frac{B^{(1)}(t)}{B_1} - B_2 \right) \right] \right)} \times \frac{Z \otimes Z - Z \otimes \I - \I \otimes Z}{\left(1 + \exp \left[-\left(\frac{B^{(2)}(t)}{B_1} - B_2 \right) \right] \right)} ,
\end{align}
with two Lindblad operators $L_1 = \frac{1}{2 \sqrt{T_1}}(X + i Y) \otimes \I$ and $L_2 = \frac{1}{2 \sqrt{T_1}} \I \otimes (X + i Y)$. Any parameters or Hamiltonians denoted by superscript $(i)$ mark either the first ($i=1$) or the second ($i=2$) qubit, and are identical to the Hamiltonian in line (\ref{eq:QDHam}). The stochastic parts for single-qubit Hamiltonians on different qubits are taken to be independent. The two-qubit control parameter $C(t)$ can only take two values, $C(t) \in \{0, C_\text{max} \}$, and the noise parameter $\gamma(t)$ is an independent zero-mean stationary Gaussian process with autocorrelation function
\begin{align}
	\langle \gamma(t_1)\gamma(t_2) \rangle &= \Gamma_{\gamma}^2 \delta(|t_1-t_2|).
\end{align}

\subsection{Physical Model 2}
Physical Model 2 is an archetypal two level system. For a single qubit, the Hamiltonian is given by
\begin{align}
	H(t) = \frac{1}{2}[B(t)(1 + \beta_1(t)) + \beta_2(t)]Z + \frac{1}{2} A(t) (1 + \alpha(t))\left[\cos(\phi(t)) X + \sin(\phi(t)) Y\right],\label{eq:SCHam}
\end{align}
with the only Lindblad operator given by $L = \frac{1}{2 \sqrt{T_1}}(X + i Y)$.

The parameters $A(t)$, $B(t)$, $\phi(t)$, and $T_1$, define the deterministic evolution of the system, with the first three representing the single-qubit controls. Every control value is specified for intervals of length $\delta t$, over which the controls remain constant. Each control scalar is bounded by some maximum value; that is $\left| A(t) \right| \leq A_\text{max}$ and $\left|B(t) \right| \leq B_\text{max}$ for some $A_\text{max},B_\text{max} \geq 0$, but is not limited by any control rates.

The parameters specified using the letters $\alpha$ and $\beta$ are all stationary Gaussian processes. All are zero-mean and independent. That is, $\langle \alpha(t) \rangle = \langle \beta_i(t) \rangle = 0$ for $i =1,2$, and $ \langle \alpha(t_1) \beta_i(t_2) \rangle = \langle \beta_1(t_1) \beta_2(t_2) \rangle = 0$ for $i =1,2$. The auto-correlation functions are given as
\begin{align}
	\langle \alpha(t_1)\alpha(t_2) \rangle &= \Gamma_{\alpha}^2~g_{1/f}(\Lambda_{\alpha}^{(l)}, \Lambda_{\alpha}^{(u)}, | t_1 - t_2|)\\
	\langle \beta_1(t_1)\beta_1(t_2) \rangle &=  \Gamma_{\beta_1}^2~g_{1/f}(\Lambda_{\beta_1}^{(l)}, \Lambda_{\beta_1}^{(u)}, | t_1 - t_2|)\\
	\langle \beta_2(t_1)\beta_2(t_2) \rangle &= \Gamma_{\beta_2}^2~g_{1/f}(\Lambda_{\beta_2}^{(l)}, \Lambda_{\beta_2}^{(u)}, | t_1 - t_2|),
\end{align}
where the parameters labelled with the letter $\Gamma$ are the noise strengths and those labelled with $\Lambda$ represent upper and lower cutoffs for $1/f$ noise. The autocorrelation function $g_{1/f}$ for $1/f$ noise is defined as the Fourier transform of $1/f$ spectral density with smooth cutoffs \cite{hooge_correlation_1997}
\begin{equation}
  \begin{aligned}
    \!g_{1/f}(\Lambda_1, \Lambda_2, \Delta t) = \int_{-\infty}^{\infty} \frac{2}{\pi \omega} \left( \arctan \left( \frac{\omega}{\Lambda_1} \right) - \arctan \left( \frac{\omega}{\Lambda_2} \right) \right) e^{-i \omega \Delta t}\, \dd\omega.
  \end{aligned}
\end{equation}
Notice that $\underset{\Lambda_1 \rightarrow 0, \Lambda_2 \rightarrow \infty}{\lim} \frac{2}{\pi \omega} \left( \arctan \left( \frac{\omega}{\Lambda_1} \right) - \arctan \left( \frac{\omega}{\Lambda_2} \right) \right) = \frac{1}{|\omega|}$.

The two-qubit Hamiltonian for this model is given by
\begin{align}
\begin{aligned}   
   H(t) &= H^{(1)}(t) \otimes \I + \I \otimes H^{(2)}(t) + H_{zz}(t) \\
   H_{zz}(t) &= -\frac{1}{2} C(t)(1+\gamma(t)) Z \otimes Z ,
\end{aligned}
\end{align}
with two Lindblad operators $L_1 = \frac{1}{2 \sqrt{T_1}}(X + i Y) \otimes \I$ and $L_2 = \frac{1}{2 \sqrt{T_1}} \I \otimes (X + i Y)$.

Single-qubit Hamiltonians denoted by $H^{(i)}$ acting either on the first ($i=1$) or the second ($i=2$) qubit have identical parameters to the Hamiltonian in Equation (\ref{eq:SCHam}), and stochastic parts for single-qubit Hamiltonians are taken to be independent. The two-qubit control parameter $C(t)$ is bounded in its maximum value $|C(t)| \leq C_\text{max}$, but is otherwise unconstrained. $\gamma(t)$ is an independent zero-mean stationary Gaussian process, its autocorrelation function being given by
\begin{align}
	\langle \gamma(t_1)\gamma(t_2) \rangle &= \Gamma_{\gamma}^2~g_{1/f}(\Lambda_{\gamma}^{(l)}, \Lambda_{\gamma}^{(u)}, | t_1 - t_2|) .
\end{align}

\subsection{XY Sequence Gate Protocol} \label{app:pulse-protocol}
Suppose we have a dynamical decoupling sequence which is given as a list of unitary operations $\{ A_i\}$, $i=1,...,N$, where $i$ denotes the temporal order of these operations. We demand that
\begin{equation}
\label{eq:decouplingCondition}
A_N~...~A_2A_1 = e^{i \phi}~\I ,
\end{equation}
where $e^{i\phi}$ is an arbitrary global phase. If we want to spread a unitary gate $U$ across the sequence $\{ A_i\}$, we first find
\begin{align}
\label{eq:unitaryBits}
U_1 &= A_1U^{1/N}A_1^{-1} \\
U_2 &= A_2A_1U^{1/N}A_1^{-1}A_2^{-1} \\
&\vdots \\
U_N &= A_N~...~A_2A_1U^{1/N}A_1^{-1}A_2^{-1}~...~A_N^{-1},
\end{align}
where $\left( U^{1/N} \right)^N = U$, and then implement the sequence $A_1$, $U_1$, $A_2$, $U_2$,..., $A_N$, $U_N$ resulting in $U_NA_N~...~U_2A_2U_1A_1 = e^{i\phi}U$, which follows from direct substitution and Equation (\ref{eq:decouplingCondition}).

\subsubsection{XY8 sequence}
The XY8 sequence \cite{gullion_new_1990} is an $8$-unitary decoupling sequence where, following the notation above, $A_1 = A_3 = A_6 = A_8 = X$ and $A_2 = A_4 = A_5 = A_7 = Y$. In the limit of perfect control and infinitesimally short (delta) pulses, the sequence refocuses noise along any direction that varies slower than the sequence is implemented. To implement a unitary gate $U$ within the sequence, we split it into 8 parts,
\begin{equation}
  \begin{array}{c}
	U_1 = U_7 = XU^\frac{1}{8}X,\quad
	U_2 = U_6 = YXU^\frac{1}{8}XY,\\
	U_3 = U_5 = XYXU^\frac{1}{8}XYX,\quad
	U_4 = U_8 = U^\frac{1}{8} ,
  \end{array}
\end{equation}
where we simplify the expression using $XYXY=YXYX=-\I$, and the fact that the global phase of the desired unitary is irrelevant.

\subsubsection{XY4 sequence}
The XY4 sequence \cite{maudsley_modified_1986} is a $4$-unitary decoupling sequence where, following the notation above, $A_1 = A_3 = X$ and $A_2 = A_4 = Y$. Like the XY8 sequence, the XY4 sequence refocuses noise along any direction that varies slower than the sequence is implemented, given that the pulses are ideal and infinitesimally short. We spread a unitary gate $U$ across the sequence by breaking it into four parts,
\begin{equation}
  \begin{array}{c}
    U_1 = XU^\frac{1}{4}X,\quad
	U_2 = YXU^\frac{1}{4}XY,\\
	U_3 = XYXU^\frac{1}{4}XYX,\quad
	U_4 = U^\frac{1}{4} .
  \end{array}
\end{equation}

\subsection{Gate Sets}
Gate Set 1 (GS1) is built on Physical Model 1 using the parameters in Table \ref{tab:gs1-params}. This gate set was built from an XY8 pulse sequence \cite{gullion_new_1990}, with single-qubit gates being implemented within this sequence according to the XY sequence gate protocol. Each pulse piece in the sequence was found via the GRAPE algorithm \cite{khaneja_optimal_2005,borneman_application_2010} with control constraints from Table \ref{tab:gs1-params} incorporated into the algorithm. All gates are $199.2~\text{ns}$ long and the discretization step for cumulant simulations was chosen to be $0.1~\text{ns}$.

Gate set 2 (GS2) is built on Physical Model 2 using the parameters in Table \ref{tab:gs2-params}. This gate set was built from an XY4 pulse sequence \cite{maudsley_modified_1986}, again, with single-qubit gates being implemented within this sequence according to the XY sequence gate protocol. All pulse pieces are performed using hard pulses. All gates are $168~\text{ns}$ long and the discretization step for cumulant simulations was chosen to be\IfJournal{ $0.25~\text{ns}$. (As GS2 uses hard pulses, the cumulant simulation can be discretized more coarsely, as the pulse amplitudes and phases remain constant for longer periods of time. The same can be said for GS3, though given that the gates are so short, a smaller time step was used anyway.)}\IfArxiv{$0.25~\text{ns}$\footnote{As GS2 uses hard pulses, the cumulant simulation can be discretized more coarsely, as the pulse amplitudes and phases remain constant for longer periods of time. The same can be said for GS3, though given that the gates are so short, a smaller time step was used anyway.}.}

Gate set 3 (GS3) is also built on Physical Model 2, but uses the noise parameters in Table \ref{tab:gs3-params} to provide variety in the resultant gate errors. No refocusing pulse sequences are used; all gates are generated from simple hard pulses. All gates are $25~\text{ns}$ long and the discretization step for cumulant simulations was chosen to be $0.1~\text{ns}$.

For each gate set, the two-qubit \cnotgt~gate is implemented using the identity
\begin{align}
	(-1)^\frac{3}{4}U_{\text{\cnotgt~}} = e^{i \frac{\pi}{2} \I \otimes \frac{X}{2}}e^{-i\frac{\pi}{2} \I \otimes \frac{Y}{2}}e^{-i \pi \frac{Z \otimes Z}{4}}e^{i \frac{\pi}{2} \I \otimes \frac{Y}{2}} e^{i \frac{\pi}{2} \frac{Z}{2} \otimes \I}.\label{eqn:cnot-ident}
\end{align}
For GS1 and GS2, as for single-qubit gates, the gate is broken into parts that are interspersed into their respective XY sequences. In this case however, the first two single-qubit rotations are done during the first half of the XY sequence, and the last two single-qubit rotations are done during the second half of the XY sequence, in a way similar to the single-qubit gates. The two-qubit coupling operation is implemented in the middle of the XY sequence. For GS3, the \cnotgt~gate is implemented according to the above identity, using hard pulses.

See Appendix \ref{app:cumulant} for details on the procedure used to simulate the gates.

\begin{table}[t!] 
\centering 
\caption{\label{tab:gs1-params} Parameters used for Physical Model 1, Gate Set 1.}
\begin{tabular}{lrllrl}
\cmidrule[\heavyrulewidth](r){1-3}\cmidrule[\heavyrulewidth](l){4-6}
Control & \multicolumn{2}{r}{\multirow{2}{*}{Value}} & Noise &\multicolumn{2}{r}{\multirow{2}{*}{Value}} \\
Parameter & & & Parameter & & \\
\cmidrule(r){1-3}\cmidrule(l){4-6}
$B_0$                 & $1.5193 \times 10^{13}$  & Hz     & $T_1$               & $1$                & s  \\
$B_1$                 & $1.5193 \times 10^{11}$  & Hz     & $\Gamma_{\alpha_1}$ & $4.804$            & Hz \\
$B_2$                 & $120$                   &        & $\Gamma_{\alpha_2}$ & $1.519 \times 10^8$ & Hz \\
$A_\text{max}$        & $3.798 \times 10^8$      & Hz     & $\tau_1$            & $10^{-2}$          & s  \\
$B_\text{max}$        & $3.0385 \times 10^{13}$  & Hz     & $\tau_2$            & $10^{-3}$          & s  \\
$\Delta A_\text{max}$ & $0.7596 \times 10^{18}$  & Hz / s & $\tau_3$            & $10^{-4}$          & s  \\
$\Delta B_\text{max}$ & $1.215 \times 10^{23}$   & Hz / s & $\Gamma_{\beta_1}$  & $1.519 \times 10^9$ & Hz \\
$C_\text{max}$        & $8.73568 \times 10^{12}$ & Hz     & $\Gamma_{\beta_2}$  & $4.804 \times 10^6$ & Hz \\
$\delta t$            & $10^{-10}$              & s      & $\Gamma_{\gamma}$   & $10^3$             & Hz \\
\cmidrule[\heavyrulewidth](r){1-3}\cmidrule[\heavyrulewidth](l){4-6}
\end{tabular}
\end{table}

\begin{table}[t!] 
\centering 
\caption{\label{tab:gs2-params} Parameters used for Physical Model 2, Gate Set 2}
\begin{tabular}{lrllrl}
\cmidrule[\heavyrulewidth](r){1-3}\cmidrule[\heavyrulewidth](l){4-6}
Control & \multicolumn{2}{r}{\multirow{2}{*}{Value}} & Noise &\multicolumn{2}{r}{\multirow{2}{*}{Value}} \\
Parameter & & & Parameter & & \\
\cmidrule(r){1-3}\cmidrule(l){4-6}
$A_\text{max}$ & $2 \pi \times 10^8$ & Hz & $T_1$                     & $10^{-4}$                & s  \\ 
$B_\text{max}$ & $2 \pi \times 10^9$ & Hz & $\Gamma_{\alpha}$         & $3 \times 10^4$           & Hz \\
$C_\text{max}$ & $2 \pi \times 10^8$ & Hz & $\Gamma_{\beta_1}$        & $3 \times 10^4$           & Hz \\
$\delta t$     & $10^{-9}$          & s  & $\Gamma_{\beta_2}$        & $10^6 / 2 \pi$           & Hz \\
\cmidrule[\heavyrulewidth](r){1-3}
               &                    &    & $\Lambda_{\alpha}^{(l)}$  & $1 / 2 \pi$              & Hz \\
               &                    &    & $\Lambda_{\alpha}^{(u)}$  & $10^9$                   & Hz \\
               &                    &    & $\Lambda_{\beta_1}^{(l)}$ & $1 /2 \pi$               & Hz \\
               &                    &    & $\Lambda_{\beta_1}^{(u)}$ & $10^9$                   & Hz \\
               &                    &    & $\Lambda_{\beta_2}^{(l)}$ & $1 / 2 \pi$              & Hz \\
               &                    &    & $\Lambda_{\beta_2}^{(u)}$ & $10^9$                   & Hz \\
               &                    &    & $\Gamma_{\gamma}$         & $1.2 \times 10^3 / 2 \pi$ & Hz \\
               &                    &    & $\Lambda_{\gamma}^{(l)}$  & $1 / 2 \pi$              & Hz \\
               &                    &    & $\Lambda_{\gamma}^{(u)}$  & $10^9$                   & Hz \\
\cmidrule[\heavyrulewidth](l){4-6}
\end{tabular}
\end{table}

\begin{table} 
\centering 
\caption{\label{tab:gs3-params} Parameters used for Physical Model 2, Gate Set 3}
\begin{tabular}{lrllrl}
\cmidrule[\heavyrulewidth](r){1-3}\cmidrule[\heavyrulewidth](l){4-6}
Control & \multicolumn{2}{r}{\multirow{2}{*}{Value}} & Noise &\multicolumn{2}{r}{\multirow{2}{*}{Value}} \\
Parameter & & & Parameter & & \\
\cmidrule(r){1-3}\cmidrule(l){4-6}
$A_\text{max}$ & $2 \pi \times 10^8$ & Hz & $T_1$                     & $10^{-5}$                & s  \\
$B_\text{max}$ & $2 \pi \times 10^9$ & Hz & $\Gamma_{\alpha}$         & $0$                      & Hz \\
$C_\text{max}$ & $2 \pi \times 10^8$ & Hz & $\Gamma_{\beta_1}$        & $10^4$                   & Hz \\
$\delta t$     & $10^{-9}$          & s  & $\Gamma_{\beta_2}$        & $10^4$                   & Hz \\
\cmidrule[\heavyrulewidth](r){1-3}
               &                    &    & $\Lambda_{\alpha}^{(l)}$  & $1 / 2 \pi$              & Hz \\
               &                    &    & $\Lambda_{\alpha}^{(u)}$  & $10^9$                   & Hz \\
               &                    &    & $\Lambda_{\beta_1}^{(l)}$ & $1 / 2 \pi$              & Hz \\
               &                    &    & $\Lambda_{\beta_1}^{(u)}$ & $10^9$                   & Hz \\
               &                    &    & $\Lambda_{\beta_2}^{(l)}$ & $1 / 2 \pi$              & Hz \\
               &                    &    & $\Lambda_{\beta_2}^{(u)}$ & $10^9$                   & Hz \\
               &                    &    & $\Gamma_{\gamma}$         & $1.2 \times 10^3 / 2 \pi$ & Hz \\
               &                    &    & $\Lambda_{\gamma}^{(l)}$  & $1 / 2 \pi$              & Hz \\
               &                    &    & $\Lambda_{\gamma}^{(u)}$  & $10^9$                   & Hz \\
\cmidrule[\heavyrulewidth](l){4-6}
\end{tabular}
\end{table}

\section{Secondary Analysis}
\label{app:secondaryanalysis}

This appendix contains analysis that is secondary to the main point of the paper, but that we consider important in its own right. We identify the good and poor regimes of performance for the ``Pauli twirled'' errors, and classify our gate sets into these regimes. This analysis also aids in the understanding of the behaviour of the approximations generated by our own method; in particular the observation made in the main body that in some cases, the average fidelity of our approximations is much less than that of the original, whereas in some cases they are very similar. Given the importance on the classification of errors to this discussion, we conclude by explaining why the errors for each gate set take the form that they do.

The identification of these regimes requires analysis on what happens to the IO distinguishability properties of a channel when it is twirled. Generally, twirling a map $\Lambda$ by a set of unitaries $\{U_k\}_{k=1}^N$ is the action of mapping $\Lambda \rightarrow \frac{1}{N} \sum_{k=1}^N U^\dagger_k \circ \Lambda \circ U_k$. The twirled map results from choosing a unitary operator from the twirling set with uniform probability, applying it, applying $\Lambda$, then inverting the twirling operator. If the twirling set is chosen to be the Pauli operators, it is called a Pauli twirl and, if perfectly implemented, will transform any map into the Pauli channel that results from mathematically truncating the off-diagonal elements of the process ($\chi$-)matrix \cite{chuang_prescription_1997}.

Before analyzing the effects of Pauli twirling specifically, we can look at the general effect of twirling on the diamond norm distance of an arbitrary channel to the identity operation. Let $\H$ denote a finite dimensional Hilbert space, and $L(\H)$ the set of linear operators from $\H \rightarrow \H$. One property of the diamond norm is that for $\Phi: L(\H) \rightarrow L(\H)$, and any unitary operators $U,V \in L(\H)$, it holds that $\| U \circ \Phi \circ V \|_\diamond = \| \Phi \|_\diamond$ \cite{watrous_course_notes_2011}. Thus, for any finite set of unitaries $\{ U_k \}_{k=1}^N \subset L(\H)$, and quantum channel $\Lambda: L(\H) \rightarrow L(\H)$, it holds by straightforward application of the triangle inequality that
\begin{align}
	\left\| \I_{L(\H)} - \frac{1}{N}\sum_{k=1}^N U^\dagger_k \circ \Lambda \circ U_k \right\|_\diamond &\leq \left\| \I_{L(\H)} - \Lambda \right\|_\diamond,
\end{align}
where $\I_{L(\H)}$ is the identity channel. In other words, the distance of a twirled error to the identity operation is always bounded above by that of the original error and so, in a worst case sense, twirling typically acts to make an error harder to detect.

To specify the regimes of performance of twirling, we look at the Bloch representation of quantum channels. Any qubit map $\Lambda$ can be represented as a matrix $M$ and vector $\vec{t}$ that acts on Bloch vectors as $\vec{r} \rightarrow M \vec{r} + \vec{t}$. $M$ can be written in terms of its polar decomposition $M = OP$, where $O$ and $P$ are an orthogonal and positive semidefinite matrix, respectively. Thus, the action on the Bloch sphere can be represented as $\vec{r} \rightarrow O( P \vec{r} + O^T \vec{t})$ \cite{Nielsen2004Quantum}. That is, as a possibly non-unital channel followed by an orthogonal rotation of the Bloch sphere, which corresponds to a unitary rotation for qubits \cite{KingRuskai01}. We say that a channel is in the ``unitary regime'' if the effect of $O$ is relatively large compared to $P$ and $\vec{t}$, and we say the channel is in the ``deforming regime'' if the opposite is true. To quantify the ``effect'' of a Bloch matrix $M$, we use the quantity $\| \I - M \|_2$, and use $\| \vec{t} \|_2$ to quantify the effect of the non-unital part. We use $\| \cdot \|_2$ to denote both the Hilbert-Schmidt norm on matrices and the Euclidean norm on vectors, where context and notation will make clear which is meant. In the following paragraphs, we examine how twirling affects worst-case IO distinguishability properties of different errors in these regimes.

First, for a unital deforming qubit channel $\Lambda$, the Bloch representation is simply a positive semi-definite matrix $P$. Diagonalize $P$ as $P=U D U^\dagger$, for some orthogonal matrix $U$ and diagonal non-negative matrix $D$. Then, as diagonal Bloch matrices are realizable as Pauli channels, we can write $\Lambda$ as $\Lambda(\rho) =  \sum_{i=0}^3 p_i  V P_i V^\dagger \rho  V P_i  V^\dagger$, where $V$ is the unitary corresponding to $U$. Channels of this form are called Generalized Pauli channels, and the diamond norm distance between channels of this form with different probability vectors $\vec{p}$ and $\vec{q}$ is given by $\sum_{i=0}^3 |p_i - q_i|$ \cite{sacchi_minimum_2005, magesan_characterizing_2012}. For these channels, as $V P_i V^\dagger$ has no identity part for $i \geq 1$, the $\chi_{00}$ element in the Pauli basis is identical to the probability assigned to the identity operator. As this quantity is conserved in Pauli twirling, and as the identity map on qubits $\I_{L(\complex^2)}$ is a Generalized Pauli channel, it follows that, for a deforming unital qubit channel $\Lambda$ and its Pauli twirl $\Lambda_\text{PT}$, $\| \I_{L(\complex^2)} - \Lambda \|_\diamond = (1-p_0) + p_1 + p_2 + p_3 = 2(1-p_0) = \| \I_{L(\complex^2)} -\Lambda_\text{PT} \|_\diamond$. Thus, diamond norm distance to the identity for these channels is unaffected by Pauli twirling.

To examine the qualitative behaviour of Pauli twirling on a non-unital deforming channel, we examine the special case of an amplitude damping channel, which has Kraus operators
\begin{align}
	K_1 = \left(\begin{array}{cc} 1 & 0 \\ 0 & \sqrt{1-\gamma} \end{array} \right),\; K_2 = \left(\begin{array}{cc} 0 & \sqrt{\gamma} \\ 0 & 0 \end{array} \right),
\end{align}
for some parameter $\gamma$, and Bloch representation
\begin{align}
	M  = \left( \begin{array}{ccc} \sqrt{1-\gamma} & 0 & 0 \\ 0 & \sqrt{1-\gamma} & 0\\ 0 & 0 & 1-\gamma \end{array} \right), \;
	\vec{t} = \left(\begin{array}{c} 0\\ 0 \\ \gamma \end{array} \right).
\end{align}
In the Bloch representation, the effect of Pauli twirling is to remove the off-diagonal elements of $M$ and set $\vec{t} \rightarrow 0$. Thus, the only effect that twirling has on this channel is to remove the non-unital part. Note that the dominant error considered by Geller and Zhou in \cite{geller_efficient_2013} is of this form; the Bloch matrix is diagonal, and so in some sense, Pauli twirling has a minimal effect. We consider the worst case performance of this channel on a pure qubit state. It is clear that the state most affected by $\Lambda$, and by its Pauli twirl $\Lambda_\text{PT}$, is the $-1$ eigenstate of $Z$, which we denote as $\rho_-$. It can be easily checked that $\| \rho_- - \Lambda(\rho_-) \|_1 = 2 \gamma$, and $\| \rho_- - \Lambda_\text{PT}(\rho_-) \|_1 = \gamma$. Thus, while the worst case performance in this case is lessened by Pauli twirling, it is only by a factor of $2$.

Lastly, we look at purely unitary channels. For two unitary operators $U, V$ in $L(\H)$, there exists a pure state $\ket{\psi} \in \H$ for which 
\begin{align}
	\| U\cdot U^\dagger - V \cdot V^\dagger \|_\diamond = \|U \ket{\psi}\bra{\psi} U^\dagger - V \ket{\psi} \bra{\psi} V^\dagger \|_1 = 2\sqrt{1- |\bra{\psi}U^\dagger V \ket{\psi}|^2}
\end{align} 
(see \cite{watrous_course_notes_2011}). From this form, it is clear that any state that maximizes the distinguishability between the identity operation and a unitary $U$ will be orthogonal to the unitarys rotation axis. Thus, for a rotation $U$ by an angle $\theta$, $\| \I_{L(\complex^2)} - U \cdot U^\dagger\|_\diamond = 2|\sin(\theta/2)|$. As the $\chi_{00}$ element of a qubit rotation by angle $\theta$ is $\cos^2(\theta/2)$, it follows from the preceding discussion that, for the Pauli twirled error $\Lambda_\text{PT}$, $\| \I_{L(\complex^2)} - \Lambda_\text{PT} \|_\diamond = 2(1-\cos^2(\theta/2)) = 2 \sin^2(\theta/2)$. Thus, for small values of $\theta$, the twirled channel can be orders of magnitude closer to the identity than the original.

With this qualitative analysis in hand, we examine the form of the Bloch representation of the errors considered in this paper. The Bloch representation of each error is decomposed into the three pieces $O$, $P$, and $\vec{t}$, and the size of each piece is reported in Table \ref{table:bloch-stats}. As the identity gate occurs most frequently in the circuit, its properties are the most important.

\begin{table} 
\caption{\label{table:bloch-stats} We denote the Bloch representation of each gate error as $(OP, \vec{t})$, where $O$ is an orthogonal matrix, $P$ is positive-semidefinite, and $\vec{t}$ is the ``non-unital part''. Norms of size less than $10^{-10}$ are displayed as $0$, as at this size they are irrelevant compared to the dominant parts of the error.}
\begin{center}
\begin{tabular}{ccllllll} \toprule 
\multicolumn{2}{c}{Set/Statistics} & $\I$& $X$& $Y$& $Z$& $H$& \cnotgt \\ \midrule 
\multirow{3}{*}{GS1} &  $\| \I - O \|_2$ & $6.74\times 10^{-3}$ & $1.83\times 10^{-2}$ & $1.83\times 10^{-2}$ & $5.06\times 10^{-3}$ & $1.63\times 10^{-2}$ & $5.78\times 10^{-2}$\\ 
 &  $\| \I - P \|_2$ & $8.22\times 10^{-7}$ & $1.06\times 10^{-2}$ & $4.43\times 10^{-3}$ & $1.33\times 10^{-2}$ & $6.24\times 10^{-3}$ & $3.56\times 10^{-2}$\\ 
 & $\|\vec{t}\|_2$ & $0.$ & $5.51\times 10^{-9}$ & $5.56\times 10^{-9}$ & $0.$ & $1.36\times 10^{-8}$ & $2.19\times 10^{-8}$\\ 
\midrule
\multirow{3}{*}{GS2} &  $\| \I - O \|_2$ & $3.35\times 10^{-5}$ & $5.52\times 10^{-4}$ & $5.52\times 10^{-4}$ & $2.09\times 10^{-4}$ & $1.59\times 10^{-3}$ & $1.73\times 10^{-3}$\\ 
 &  $\| \I - P \|_2$ & $2.16\times 10^{-3}$ & $2.15\times 10^{-3}$ & $2.15\times 10^{-3}$ & $2.32\times 10^{-3}$ & $2.19\times 10^{-3}$ & $8.13\times 10^{-3}$\\ 
 & $\|\vec{t}\|_2$ & $4.01\times 10^{-8}$ & $5.65\times 10^{-5}$ & $4.95\times 10^{-5}$ & $1.66\times 10^{-4}$ & $9.91\times 10^{-5}$ & $2.14\times 10^{-4}$\\ 
\midrule
\multirow{3}{*}{GS3} &  $\| \I - O \|_2$ & $0.$ & $1.09\times 10^{-5}$ & $1.50\times 10^{-5}$ & $1.04\times 10^{-5}$ & $2.83\times 10^{-6}$ & $4.45\times 10^{-2}$\\ 
 &  $\| \I - P \|_2$ & $3.06\times 10^{-3}$ & $2.93\times 10^{-3}$ & $2.93\times 10^{-3}$ & $3.06\times 10^{-3}$ & $2.96\times 10^{-3}$ & $1.12\times 10^{-2}$\\ 
 & $\|\vec{t}\|_2$ & $2.50\times 10^{-3}$ & $1.37\times 10^{-3}$ & $1.37\times 10^{-3}$ & $2.50\times 10^{-3}$ & $2.11\times 10^{-3}$ & $2.95\times 10^{-3}$\\ 
\bottomrule  
 \end{tabular}
\end{center}
\end{table}

We see that GS1 has the largest unitary component to its errors (in terms of the ratio to the other components). The error in the identity gate is almost entirely unitary, and therefore falls neatly into the unitary regime. Indeed, this is consistent with the fact that a single-qubit unitary rotation by an angle $\theta$ will have a $\chi_{00}$ element of $\cos^2(\theta/2) \approx 0.99999$, whereas the Pauli channel that \emph{exactly} matches its IO distinguishably has a $\chi_{00}$ element of $1 - |\sin(\theta/2)| \approx 0.99762$. We see that these numbers correspond exactly to the given number of digits in Table \ref{table:identity-stats}, where a two-order-of-magnitude decrease in the distance of the Pauli twirled error to the identity channel is shown. While the other errors in the set have unitary parts, the non-unitary parts are comparable in size. As such, character from both the unitary and unital deforming regimes is observed; the diamond norm distance of the Pauli twirled error to the identity decreases, but the decrease is not as impressive as for the identity gate. For the $Y$ and $H$ errors, the unitary part is $2$ or $3$ times larger than the non-unitary, and thus the decrease in diamond norm distance to the ideal channel for the Pauli twirled errors is the greatest of the non-identity gates.

For every gate in GS2, $\| \vec{t} \|_2$ is relatively small and $\| \I - O \|_2$ is in most cases an order of magnitude smaller than $\| \I - P \|_2$, putting GS2 into the unital deforming regime. When looking at the diamond norm distances of the Pauli twirled channels to the identity, we see that, as expected in this regime, there is very little decrease. Indeed, while the Pauli twirled channels generally underestimate the error in the hedging metrics, the underestimation is small.

For GS3, $\| \I - O \|_2$ is usually of comparatively negligible size, and $\| \vec{t} \|_2$ is of the same order of magnitude as $\| \I - P \|_2$, putting this gate set into the non-unital deforming regime. Looking at the decrease of the diamond norm distance to the ideal operations for the Pauli twirled errors, we see roughly what is expected; a decrease on the order of a factor of $2$. As in the case of GS2, while the Pauli twirled errors show underestimation in the hedging statistics, it is by a small degree.

This discussion supports and illuminates the idea of two regimes of behaviour for Pauli twirled approximations. In the unitary case, we observe large underestimation of IO distinguishability properties. For the circuit we've considered, this underestimation propagates upwards; the gadget simulated with Pauli twirled errors is much harder to distinguish from the perfect gadget than the original errors. In the deformation regime, the Pauli twirled errors can result in underestimation. One might argue however, as Geller and Zhou do, that this level of underestimation is acceptable. Indeed, while the hedging is negative for the Pauli twirled gates and circuit for GS2, it is small in magnitude. For GS3, the Pauli twirled circuit even has positive hedging. This is perhaps not surprising on an intuitive level. Pauli channels are a subset of deforming channels, and while not all deforming errors are Pauli channels, they are \emph{like} Pauli channels. All deforming channels contract the Bloch sphere inwards (with possible non-unital shifts), so Pauli twirling simply maps one deforming channel to another, which doesn't dramatically change the IO distinguishability properties. This stands in contrast to the unitary regime where, in regards to IO distinguishability, something fundamental can be lost when a dominantly unitary error is mapped to a Pauli channel via Pauli twirling.

This discussion also demonstrates why our approximations behave the way they do in regards to average gate fidelity. As we've seen from the Pauli twirl analysis, for a fixed average fidelity, channels in the unitary regime are much more distinguishable from the identity operation than those in the deforming regime. Thus, if the original error is in the unitary regime, our approximations, which fall into the deforming regime, will sometimes have a much worse average fidelity than the original to allow them to honestly represent its IO distinguishability properties.

As a final point, we can ask why the errors for each gate set have the form that they do. To explain this, it is necessary to examine the physical models and the pulse design techniques used, which are described in detail in Appendix \ref{app:physical-models}. The simplest case is GS3, where hard pulses with no refocusing sequences were used. As the $T_1$ time was the shortest of all of the gate sets, and no refocusing pulses were used, the strong non-unital effects were allowed to continue building in their natural direction throughout each gate resulting in primarily non-unital errors. In GS2, which used the same physical model as GS3, the $T_1$ time was an order of magnitude longer. In addition, the refocusing pulse sequence used has the effect of ``flipping'' the non-unital shift between the positive and negative $z$-direction, and so, when averaged over the whole sequence, the effective non-unital shift tends to zero. As a result, the size of the non-unital piece tends to be at least an order of magnitude smaller than the other pieces, putting this gate set into the unital deformation regime. Finally, as GS1 had the longest $T_1$ time, and the refocusing sequence was performed at a rate far faster than the $T_1$ time, the non-unital error components are negligible. Aside from this, the prime difference between GS1 and the other gate sets is that the pulses used were found using the GRAPE algorithm, rather than being hard, which accounts for the dominant unitary components of the errors. For hard pulses, in the limit of no noise, the implemented unitary should be perfect, aside from numerical imprecision in the field amplitudes, which can be made arbitrarily small with low overhead. For pulses found using the GRAPE algorithm this is not the case; even a noiseless implementation will not produce the perfect unitary. Arguably, this situation is more prevalent in experimental implementations as pulse finding algorithms like GRAPE are necessary to obtain high-fidelity control over systems with complicated Hamiltonians. In addition, these types of algorithms can be used to achieve high-fidelity control across a range of internal Hamiltonian parameters, even when the Hamiltonian is simple \cite{borneman_application_2010}.

\pagebreak
\section{Gate Statistics} \label{app:gate-statistics}

\begin{table}[H]
\caption{Statistics for the various approximations of the gates in GS1, approximated from $N=10^6$ random pure states.}
\begin{tabular}{ccllll} \toprule 
\multicolumn{6}{c}{GS1} \\ \midrule 
\multicolumn{2}{c}{Gate/Statistics} & $\hphantom{-}$Original& $\hphantom{-}$Pauli twirled& $\hphantom{-}$Pauli& $\hphantom{-}$Clifford \\ \midrule 
\multirow{6}{*}{$\I$} & $\chi_{00}$ & $\hphantom{-}0.999994$ & $\hphantom{-}0.999994$ & $\hphantom{-}0.997618$ & $\hphantom{-}0.998314$ \\ 
 			& $\| \Lambda - U_{\text{Ideal}} \|_{\diamond}$ & $\hphantom{-}4.76\times 10^{-3}$ & $\hphantom{-}1.20\times 10^{-5}$ & $\hphantom{-}4.76\times 10^{-3}$ & $\hphantom{-}4.77\times 10^{-3}$ \\ 
 			& $\| \Lambda - \Lambda_{\text{Original}} \|_{\diamond}$ &  & $\hphantom{-}4.76\times 10^{-3}$ & $\hphantom{-}6.73\times 10^{-3}$ & $\hphantom{-}3.64\times 10^{-3}$ \\ 
 			& $\overline{h}$ &  & $-3.73\times 10^{-3}$ & $\hphantom{-}1.14\times 10^{-7}$ & $\hphantom{-}1.64\times 10^{-6}$ \\ 
 			& $p_{\text{viol}}$ &  & $\hphantom{-}1.$ & $\hphantom{-}0.$ & $\hphantom{-}0.$ \\   \midrule 
\multirow{6}{*}{$X$} & $\chi_{00}$ & $\hphantom{-}0.996147$ & $\hphantom{-}0.996147$ & $\hphantom{-}0.991234$ & $\hphantom{-}0.994179$ \\ 
 			& $\| \Lambda - U_{\text{Ideal}} \|_{\diamond}$ & $\hphantom{-}1.51\times 10^{-2}$ & $\hphantom{-}7.71\times 10^{-3}$ & $\hphantom{-}1.75\times 10^{-2}$ & $\hphantom{-}1.59\times 10^{-2}$ \\ 
 			& $\| \Lambda - \Lambda_{\text{Original}} \|_{\diamond}$ &  & $\hphantom{-}1.29\times 10^{-2}$ & $\hphantom{-}1.67\times 10^{-2}$ & $\hphantom{-}4.58\times 10^{-3}$ \\ 
 			& $\overline{h}$ &  & $-5.76\times 10^{-3}$ & $\hphantom{-}1.12\times 10^{-3}$ & $\hphantom{-}3.90\times 10^{-4}$ \\ 
 			& $p_{\text{viol}}$ &  & $\hphantom{-}0.99892$ & $\hphantom{-}0.$ & $\hphantom{-}0.$ \\  \midrule 
\multirow{6}{*}{$Y$} & $\chi_{00}$ & $\hphantom{-}0.998330$ & $\hphantom{-}0.998330$ & $\hphantom{-}0.992021$ & $\hphantom{-}0.994711$ \\ 
 			& $\| \Lambda - U_{\text{Ideal}} \|_{\diamond}$ & $\hphantom{-}1.35\times 10^{-2}$ & $\hphantom{-}3.34\times 10^{-3}$ & $\hphantom{-}1.60\times 10^{-2}$ & $\hphantom{-}1.43\times 10^{-2}$ \\ 
 			& $\| \Lambda - \Lambda_{\text{Original}} \|_{\diamond}$ &  & $\hphantom{-}1.30\times 10^{-2}$ & $\hphantom{-}1.86\times 10^{-2}$ & $\hphantom{-}8.15\times 10^{-3}$ \\ 
 			& $\overline{h}$ &  & $-7.96\times 10^{-3}$ & $\hphantom{-}9.94\times 10^{-4}$ & $\hphantom{-}3.09\times 10^{-4}$ \\ 
 			& $p_{\text{viol}}$ &  & $\hphantom{-}0.99953$ & $\hphantom{-}0.$ & $\hphantom{-}0.$ \\   \midrule 
\multirow{6}{*}{$Z$} & $\chi_{00}$ & $\hphantom{-}0.995289$ & $\hphantom{-}0.995289$ & $\hphantom{-}0.992055$ & $\hphantom{-}0.993663$ \\ 
 			& $\| \Lambda - U_{\text{Ideal}} \|_{\diamond}$ & $\hphantom{-}1.06\times 10^{-2}$ & $\hphantom{-}9.42\times 10^{-3}$ & $\hphantom{-}1.59\times 10^{-2}$ & $\hphantom{-}1.31\times 10^{-2}$ \\ 
 			& $\| \Lambda - \Lambda_{\text{Original}} \|_{\diamond}$ &  & $\hphantom{-}4.59\times 10^{-3}$ & $\hphantom{-}7.68\times 10^{-3}$ & $\hphantom{-}3.46\times 10^{-3}$ \\ 
 			& $\overline{h}$ &  & $-5.71\times 10^{-4}$ & $\hphantom{-}3.01\times 10^{-3}$ & $\hphantom{-}1.29\times 10^{-3}$ \\ 
 			& $p_{\text{viol}}$ &  & $\hphantom{-}0.60369$ & $\hphantom{-}0.$ & $\hphantom{-}0.$ \\  \midrule 
\multirow{6}{*}{$H$} & $\chi_{00}$ & $\hphantom{-}0.997642$ & $\hphantom{-}0.997642$ & $\hphantom{-}0.991031$ & $\hphantom{-}0.993638$ \\ 
 			& $\| \Lambda - U_{\text{Ideal}} \|_{\diamond}$ & $\hphantom{-}1.35\times 10^{-2}$ & $\hphantom{-}4.72\times 10^{-3}$ & $\hphantom{-}1.79\times 10^{-2}$ & $\hphantom{-}1.54\times 10^{-2}$ \\ 
 			& $\| \Lambda - \Lambda_{\text{Original}} \|_{\diamond}$ &  & $\hphantom{-}1.27\times 10^{-2}$ & $\hphantom{-}1.85\times 10^{-2}$ & $\hphantom{-}8.83\times 10^{-3}$ \\ 
 			& $\overline{h}$ &  & $-6.42\times 10^{-3}$ & $\hphantom{-}2.51\times 10^{-3}$ & $\hphantom{-}1.10\times 10^{-3}$ \\ 
 			& $p_{\text{viol}}$ &  & $\hphantom{-}1.$ & $\hphantom{-}0.$ & $\hphantom{-}0.$ \\ \midrule 
\multirow{6}{*}{\cnotgt} & $\chi_{00}$ & $\hphantom{-}0.993152$ & $\hphantom{-}0.993152$ & $\hphantom{-}0.976600$ & $\hphantom{-}0.976600$ \\ 
 			& $\| \Lambda - U_{\text{Ideal}} \|_{\diamond}$ & $\hphantom{-}3.09\times 10^{-2}$ & $\hphantom{-}1.37\times 10^{-2}$ & $\hphantom{-}4.68\times 10^{-2}$ & $\hphantom{-}4.68\times 10^{-2}$ \\ 
 			& $\| \Lambda - \Lambda_{\text{Original}} \|_{\diamond}$ &  & $\hphantom{-}2.82\times 10^{-2}$ & $\hphantom{-}4.52\times 10^{-2}$ & $\hphantom{-}4.52\times 10^{-2}$ \\ 
 			& $\overline{h}$ &  & $-1.17\times 10^{-2}$ & $\hphantom{-}1.52\times 10^{-2}$ & $\hphantom{-}1.52\times 10^{-2}$ \\ 
 			& $p_{\text{viol}}$ &  & $\hphantom{-}1.$ & $\hphantom{-}0.$ & $\hphantom{-}0.$ \\ \bottomrule 
\end{tabular}
\end{table}

\begin{table}[H]
\caption{Statistics for the various approximations of the gates in GS2, approximated from $N=10^6$ random pure states.}
\begin{tabular}{ccllll} \toprule 
\multicolumn{6}{c}{GS2} \\ \midrule 
\multicolumn{2}{c}{Gate/Statistics} & $\hphantom{-}$Original& $\hphantom{-}$Pauli twirled& $\hphantom{-}$Pauli& $\hphantom{-}$Clifford \\ \midrule 
\multirow{6}{*}{$\I$} & $\chi_{00}$ & $\hphantom{-}0.999087$ & $\hphantom{-}0.999087$ & $\hphantom{-}0.999085$ & $\hphantom{-}0.999086$ \\ 
 			& $\| \Lambda - U_{\text{Ideal}} \|_{\diamond}$ & $\hphantom{-}1.83\times 10^{-3}$ & $\hphantom{-}1.83\times 10^{-3}$ & $\hphantom{-}1.83\times 10^{-3}$ & $\hphantom{-}1.83\times 10^{-3}$ \\ 
 			& $\| \Lambda - \Lambda_{\text{Original}} \|_{\diamond}$ &  & $\hphantom{-}2.48\times 10^{-5}$ & $\hphantom{-}2.50\times 10^{-5}$ & $\hphantom{-}2.06\times 10^{-6}$ \\ 
 			& $\overline{h}$ &  & $-1.63\times 10^{-7}$ & $\hphantom{-}2.65\times 10^{-6}$ & $\hphantom{-}8.34\times 10^{-7}$ \\ 
 			& $p_{\text{viol}}$ &  & $\hphantom{-}0.49861$ & $\hphantom{-}0.$ & $\hphantom{-}0.$ \\ \midrule 
\multirow{6}{*}{$X$} & $\chi_{00}$ & $\hphantom{-}0.999074$ & $\hphantom{-}0.999074$ & $\hphantom{-}0.998972$ & $\hphantom{-}0.999029$ \\ 
 			& $\| \Lambda - U_{\text{Ideal}} \|_{\diamond}$ & $\hphantom{-}1.91\times 10^{-3}$ & $\hphantom{-}1.85\times 10^{-3}$ & $\hphantom{-}2.06\times 10^{-3}$ & $\hphantom{-}2.00\times 10^{-3}$ \\ 
 			& $\| \Lambda - \Lambda_{\text{Original}} \|_{\diamond}$ &  & $\hphantom{-}4.38\times 10^{-4}$ & $\hphantom{-}4.85\times 10^{-4}$ & $\hphantom{-}1.16\times 10^{-4}$ \\ 
 			& $\overline{h}$ &  & $-4.04\times 10^{-5}$ & $\hphantom{-}9.61\times 10^{-5}$ & $\hphantom{-}6.12\times 10^{-5}$ \\ 
 			& $p_{\text{viol}}$ &  & $\hphantom{-}0.7897$ & $\hphantom{-}0.$ & $\hphantom{-}0.$ \\ \midrule 
\multirow{6}{*}{$Y$} & $\chi_{00}$ & $\hphantom{-}0.999075$ & $\hphantom{-}0.999075$ & $\hphantom{-}0.998990$ & $\hphantom{-}0.999033$ \\ 
 			& $\| \Lambda - U_{\text{Ideal}} \|_{\diamond}$ & $\hphantom{-}1.91\times 10^{-3}$ & $\hphantom{-}1.85\times 10^{-3}$ & $\hphantom{-}2.02\times 10^{-3}$ & $\hphantom{-}1.99\times 10^{-3}$ \\ 
 			& $\| \Lambda - \Lambda_{\text{Original}} \|_{\diamond}$ &  & $\hphantom{-}4.32\times 10^{-4}$ & $\hphantom{-}4.72\times 10^{-4}$ & $\hphantom{-}1.03\times 10^{-4}$ \\ 
 			& $\overline{h}$ &  & $-4.02\times 10^{-5}$ & $\hphantom{-}7.32\times 10^{-5}$ & $\hphantom{-}5.58\times 10^{-5}$ \\ 
 			& $p_{\text{viol}}$ &  & $\hphantom{-}0.8674$ & $\hphantom{-}0.$ & $\hphantom{-}0.$ \\ \midrule 
\multirow{6}{*}{$Z$} & $\chi_{00}$ & $\hphantom{-}0.999023$ & $\hphantom{-}0.999023$ & $\hphantom{-}0.998869$ & $\hphantom{-}0.998901$ \\ 
 			& $\| \Lambda - U_{\text{Ideal}} \|_{\diamond}$ & $\hphantom{-}1.97\times 10^{-3}$ & $\hphantom{-}1.95\times 10^{-3}$ & $\hphantom{-}2.26\times 10^{-3}$ & $\hphantom{-}2.21\times 10^{-3}$ \\ 
 			& $\| \Lambda - \Lambda_{\text{Original}} \|_{\diamond}$ &  & $\hphantom{-}2.67\times 10^{-4}$ & $\hphantom{-}4.27\times 10^{-4}$ & $\hphantom{-}3.33\times 10^{-4}$ \\ 
 			& $\overline{h}$ &  & $-1.37\times 10^{-5}$ & $\hphantom{-}1.87\times 10^{-4}$ & $\hphantom{-}1.52\times 10^{-4}$ \\ 
 			& $p_{\text{viol}}$ &  & $\hphantom{-}0.61392$ & $\hphantom{-}0.$ & $\hphantom{-}0.$ \\ \midrule 
\multirow{6}{*}{$H$} & $\chi_{00}$ & $\hphantom{-}0.999060$ & $\hphantom{-}0.999060$ & $\hphantom{-}0.998684$ & $\hphantom{-}0.998878$ \\ 
 			& $\| \Lambda - U_{\text{Ideal}} \|_{\diamond}$ & $\hphantom{-}2.33\times 10^{-3}$ & $\hphantom{-}1.88\times 10^{-3}$ & $\hphantom{-}2.63\times 10^{-3}$ & $\hphantom{-}2.53\times 10^{-3}$ \\ 
 			& $\| \Lambda - \Lambda_{\text{Original}} \|_{\diamond}$ &  & $\hphantom{-}1.25\times 10^{-3}$ & $\hphantom{-}1.45\times 10^{-3}$ & $\hphantom{-}4.35\times 10^{-4}$ \\ 
 			& $\overline{h}$ &  & $-3.05\times 10^{-4}$ & $\hphantom{-}1.98\times 10^{-4}$ & $\hphantom{-}1.32\times 10^{-4}$ \\ 
 			& $p_{\text{viol}}$ &  & $\hphantom{-}1.$ & $\hphantom{-}0.$ & $\hphantom{-}0.$ \\ \midrule 
\multirow{6}{*}{\cnotgt} & $\chi_{00}$ & $\hphantom{-}0.998071$ & $\hphantom{-}0.998071$ & $\hphantom{-}0.997683$ & $\hphantom{-}0.997683$ \\ 
 			& $\| \Lambda - U_{\text{Ideal}} \|_{\diamond}$ & $\hphantom{-}3.99\times 10^{-3}$ & $\hphantom{-}3.86\times 10^{-3}$ & $\hphantom{-}4.63\times 10^{-3}$ & $\hphantom{-}4.63\times 10^{-3}$ \\ 
 			& $\| \Lambda - \Lambda_{\text{Original}} \|_{\diamond}$ &  & $\hphantom{-}1.01\times 10^{-3}$ & $\hphantom{-}1.33\times 10^{-3}$ & $\hphantom{-}1.33\times 10^{-3}$ \\ 
 			& $\overline{h}$ &  & $-7.38\times 10^{-5}$ & $\hphantom{-}5.44\times 10^{-4}$ & $\hphantom{-}5.44\times 10^{-4}$ \\ 
 			& $p_{\text{viol}}$ &  & $\hphantom{-}0.76905$ & $\hphantom{-}0.$ & $\hphantom{-}0.$ \\ \bottomrule 
\end{tabular}
\end{table}

\begin{table}[H]
\caption{Statistics for the various approximations of the gates in GS3, approximated from $N=10^6$ random pure states.}
\begin{tabular}{ccllll} \toprule 
\multicolumn{6}{c}{GS3} \\ \midrule 
\multicolumn{2}{c}{Gate/Statistics} & $\hphantom{-}$Original& $\hphantom{-}$Pauli twirled& $\hphantom{-}$Pauli& $\hphantom{-}$Clifford \\ \midrule 
\multirow{6}{*}{$\I$} & $\chi_{00}$ & $\hphantom{-}0.998751$ & $\hphantom{-}0.998751$ & $\hphantom{-}0.996501$ & $\hphantom{-}0.996501$ \\ 
 			& $\| \Lambda - U_{\text{Ideal}} \|_{\diamond}$ & $\hphantom{-}4.99\times 10^{-3}$ & $\hphantom{-}2.50\times 10^{-3}$ & $\hphantom{-}7.00\times 10^{-3}$ & $\hphantom{-}7.00\times 10^{-3}$ \\ 
 			& $\| \Lambda - \Lambda_{\text{Original}} \|_{\diamond}$ &  & $\hphantom{-}2.50\times 10^{-3}$ & $\hphantom{-}5.28\times 10^{-3}$ & $\hphantom{-}5.28\times 10^{-3}$ \\ 
 			& $\overline{h}$ &  & $-1.03\times 10^{-3}$ & $\hphantom{-}1.91\times 10^{-3}$ & $\hphantom{-}1.91\times 10^{-3}$ \\ 
 			& $p_{\text{viol}}$ &  & $\hphantom{-}0.74978$ & $\hphantom{-}0.$ & $\hphantom{-}0.$ \\  \midrule 
\multirow{6}{*}{$X$} & $\chi_{00}$ & $\hphantom{-}0.998751$ & $\hphantom{-}0.998751$ & $\hphantom{-}0.997447$ & $\hphantom{-}0.997674$ \\ 
 			& $\| \Lambda - U_{\text{Ideal}} \|_{\diamond}$ & $\hphantom{-}3.17\times 10^{-3}$ & $\hphantom{-}2.50\times 10^{-3}$ & $\hphantom{-}5.11\times 10^{-3}$ & $\hphantom{-}4.65\times 10^{-3}$ \\ 
 			& $\| \Lambda - \Lambda_{\text{Original}} \|_{\diamond}$ &  & $\hphantom{-}1.37\times 10^{-3}$ & $\hphantom{-}3.12\times 10^{-3}$ & $\hphantom{-}2.75\times 10^{-3}$ \\ 
 			& $\overline{h}$ &  & $-3.64\times 10^{-4}$ & $\hphantom{-}1.39\times 10^{-3}$ & $\hphantom{-}1.06\times 10^{-3}$ \\ 
 			& $p_{\text{viol}}$ &  & $\hphantom{-}0.69216$ & $\hphantom{-}0.$ & $\hphantom{-}0.$ \\ \midrule 
\multirow{6}{*}{$Y$} & $\chi_{00}$ & $\hphantom{-}0.998751$ & $\hphantom{-}0.998751$ & $\hphantom{-}0.997669$ & $\hphantom{-}0.997669$ \\ 
 			& $\| \Lambda - U_{\text{Ideal}} \|_{\diamond}$ & $\hphantom{-}3.17\times 10^{-3}$ & $\hphantom{-}2.50\times 10^{-3}$ & $\hphantom{-}4.66\times 10^{-3}$ & $\hphantom{-}4.66\times 10^{-3}$ \\ 
 			& $\| \Lambda - \Lambda_{\text{Original}} \|_{\diamond}$ &  & $\hphantom{-}1.38\times 10^{-3}$ & $\hphantom{-}2.76\times 10^{-3}$ & $\hphantom{-}2.76\times 10^{-3}$ \\ 
 			& $\overline{h}$ &  & $-3.63\times 10^{-4}$ & $\hphantom{-}1.06\times 10^{-3}$ & $\hphantom{-}1.06\times 10^{-3}$ \\ 
 			& $p_{\text{viol}}$ &  & $\hphantom{-}0.69113$ & $\hphantom{-}0.$ & $\hphantom{-}0.$ \\ \midrule 
\multirow{6}{*}{$Z$} & $\chi_{00}$ & $\hphantom{-}0.998751$ & $\hphantom{-}0.998751$ & $\hphantom{-}0.996501$ & $\hphantom{-}0.996501$ \\ 
 			& $\| \Lambda - U_{\text{Ideal}} \|_{\diamond}$ & $\hphantom{-}4.99\times 10^{-3}$ & $\hphantom{-}2.50\times 10^{-3}$ & $\hphantom{-}7.00\times 10^{-3}$ & $\hphantom{-}7.00\times 10^{-3}$ \\ 
 			& $\| \Lambda - \Lambda_{\text{Original}} \|_{\diamond}$ &  & $\hphantom{-}2.50\times 10^{-3}$ & $\hphantom{-}5.28\times 10^{-3}$ & $\hphantom{-}5.28\times 10^{-3}$ \\ 
 			& $\overline{h}$ &  & $-1.03\times 10^{-3}$ & $\hphantom{-}1.92\times 10^{-3}$ & $\hphantom{-}1.91\times 10^{-3}$ \\ 
 			& $p_{\text{viol}}$ &  & $\hphantom{-}0.75055$ & $\hphantom{-}0.$ & $\hphantom{-}0.$ \\ \midrule 
\multirow{6}{*}{$H$} & $\chi_{00}$ & $\hphantom{-}0.998751$ & $\hphantom{-}0.998751$ & $\hphantom{-}0.996883$ & $\hphantom{-}0.996940$ \\ 
 			& $\| \Lambda - U_{\text{Ideal}} \|_{\diamond}$ & $\hphantom{-}4.31\times 10^{-3}$ & $\hphantom{-}2.50\times 10^{-3}$ & $\hphantom{-}6.23\times 10^{-3}$ & $\hphantom{-}6.12\times 10^{-3}$ \\ 
 			& $\| \Lambda - \Lambda_{\text{Original}} \|_{\diamond}$ &  & $\hphantom{-}2.27\times 10^{-3}$ & $\hphantom{-}4.43\times 10^{-3}$ & $\hphantom{-}4.36\times 10^{-3}$ \\ 
 			& $\overline{h}$ &  & $-7.92\times 10^{-4}$ & $\hphantom{-}1.68\times 10^{-3}$ & $\hphantom{-}1.60\times 10^{-3}$ \\ 
 			& $p_{\text{viol}}$ &  & $\hphantom{-}0.74018$ & $\hphantom{-}0.$ & $\hphantom{-}0.$ \\  \midrule 
\multirow{6}{*}{\cnotgt} & $\chi_{00}$ & $\hphantom{-}0.997441$ & $\hphantom{-}0.997441$ & $\hphantom{-}0.981266$ & $\hphantom{-}0.981266$ \\ 
 			& $\| \Lambda - U_{\text{Ideal}} \|_{\diamond}$ & $\hphantom{-}1.98\times 10^{-2}$ & $\hphantom{-}5.12\times 10^{-3}$ & $\hphantom{-}3.75\times 10^{-2}$ & $\hphantom{-}3.75\times 10^{-2}$ \\ 
 			& $\| \Lambda - \Lambda_{\text{Original}} \|_{\diamond}$ &  & $\hphantom{-}1.83\times 10^{-2}$ & $\hphantom{-}3.90\times 10^{-2}$ & $\hphantom{-}3.90\times 10^{-2}$ \\ 
 			& $\overline{h}$ &  & $-1.15\times 10^{-2}$ & $\hphantom{-}1.48\times 10^{-2}$ & $\hphantom{-}1.48\times 10^{-2}$ \\ 
 			& $p_{\text{viol}}$ &  & $\hphantom{-}0.99988$ & $\hphantom{-}0.$ & $\hphantom{-}0.$ \\ \bottomrule 
\end{tabular}
\end{table}

\end{document}